\documentclass[10pt]{article}

\usepackage{amsmath}
\usepackage{amssymb}

\usepackage{graphicx}

\usepackage{cite}

\usepackage{color} 

\usepackage[labelfont=bf,labelsep=period,justification=raggedright]{caption}

\makeatletter
\renewcommand{\@biblabel}[1]{\quad#1.}
\makeatother

\date{}

\renewcommand{\vec}[1]{\mathbf{#1}}

\newcommand{\wh}{\widehat}
\renewcommand{\exp}[1]{\mathrm{e}^{#1}}

\begin{document}

% Title must be 150 characters or less
\begin{flushleft}
{\Large
\textbf{Locust Dynamics: Behavioral Phase Change and Swarming}
}
% Insert Author names, affiliations and corresponding author email.
\\
Chad M. Topaz$^{1,\ast}$, 
Maria R. D'Orsogna$^{2}$, 
Leah Edelstein-Keshet$^{3}$
Andrew J. Bernoff$^{4}$
\\
\bf{1} Department of Mathematics, Statistics, and Computer Science, Macalester College, Saint Paul, Minnesota, United States of America
\\
\bf{2} Department of Mathematics, California State University at Northridge, Los Angeles, California, United States of America
\\
\bf{3} Department of Mathematics, University of British Columbia, Vancouver, British Columbia, Canada
\\
\bf{4} Department of Mathematics, Harvey Mudd College, Claremont, California, United States of America
\\
$\ast$ E-mail: ctopaz@macalester.edu
\end{flushleft}

% Please keep the abstract between 250 and 300 words
\section*{Abstract}

Locusts exhibit two interconvertible behavioral phases, solitarious and gregarious. While solitarious individuals are repelled from other locusts, gregarious insects are attracted to conspecifics and can form large aggregations such as marching hopper bands. Numerous biological experiments at the individual level have shown how crowding biases conversion towards the gregarious form. To understand the formation of marching locust hopper bands, we study phase change at the collective level, and in a quantitative framework. Specifically, we construct a partial integrodifferential equation model incorporating the interplay between phase change and spatial movement at the individual level in order to predict the dynamics of hopper band formation at the population level. Stability analysis of our model reveals conditions for an outbreak, characterized by a large scale transition to the gregarious phase. A model reduction enables quantification of the temporal dynamics of each phase, of the proportion of the population that will eventually gregarize, and of the time scale for this to occur. Numerical simulations provide descriptions of the aggregation's structure and reveal transiently traveling clumps of gregarious insects. Our predictions of aggregation and mass gregarization suggest several possible future biological experiments.

% Please keep the Author Summary between 150 and 200 words
% Use first person. PLoS ONE authors please skip this step. 
% Author Summary not valid for PLoS ONE submissions.   
\section*{Author Summary}

Locusts such as \emph{Schistocerca gregaria}, \emph{Locusta migratoria}, and \emph{Chortoceites terminifera} periodically form highly destructive plagues responsible for billions of dollars in crop losses in Africa, the Middle East, Asia, and Australia. These locusts usually exist in the so-called solitarious behavioral phase and seek isolation; gregarious individuals, however, are attracted to conspecifics. Previous experimental work has uncovered the causes of phase change in individual insects: principally, sustained exposure to sparse or crowded conditions. An open problem is to understand the intrinsic roles that phase change and social interaction play in the transition from an initially disperse, solitarious population to an aggregated, destructive marching hopper band of gregarious individuals. To this end, we construct a mathematical model that describes the interplay of phase change and spatial dynamics. Through analysis and numerical simulations, we determine a critical density threshold for gregarious band formation and quantify the collective phase change over time. We also discuss implications of our work for preventative management strategies and for possible future biological experiments.

\section*{Introduction}

Outbreaks of locusts such as \emph{Schistocerca gregaria}, \emph{Locusta migratoria}, and \emph{Chortoceites terminifera} regularly afflict vast areas of Northern Africa, the Middle East, Asia, and Australia. Depending on climate and vegetation conditions, billions of voracious locusts aggregate into destructive swarms that span areas up to a thousand square kilometers. A flying locust swarm can travel a few hundred kilometers per day, stripping most of the vegetation in its path \cite{Ken1951,Alb1967,Uva1977,Rai1989}.  A recent locust plague in West Africa (2003--2005) severely disrupted agriculture, destroying \$2.5 billion in crops destined for both subsistence and export. Despite control efforts totalling \$400 million, loss rates exceeded 50\% in certain regions \cite{Bel2005,BraDjiFay2006}. These numbers alone attest to the urgency of finding better ways to predict, manage, and control locust outbreaks.

Between outbreaks, locusts are mainly antisocial creatures who live in arid regions, laying eggs in breeding grounds lush with vegetation. Resource abundance may, on occasion, support numerous hatchings, leading to a high population density. Overcrowding at resource sites promotes transition to a social state in a self-reinforcing process.  The social locust nymphs may display mass migration behavior. Within the newly formed group, individuals cohere via sensory communication, whether visual, chemical, and/or mechanical \cite{Uva1977}. Outbreaks may be exacerbated in periods of drought, when large numbers of locusts congregate on the same breeding or feeding grounds~\cite{SpeHunWat2008,ColDesSim1998,DesColSim2000}.

Locusts are \emph{phase polyphenic:} while sharing the same genotype, individuals may display different phenotypes \cite{AppHei1999,PenSim2009} that incorporate variations in morphology \cite{Dir1953}, coloration \cite{IslRoeSim1994},
reproductive features \cite{SchAlb1999} and, significantly, behavior \cite{SimMcCHag1999, RogMatDes2003}. An individual can change from a \emph{solitarious} state (preferring isolation) to a \emph{gregarious} one (seeking conspecifics). Behavioral state is plastic \cite{Uva1977,SimMcCHag1999,PenSim2009} and strongly dependent on local population density: in sparse surroundings, a gregarious locust transitions to the solitarious state \cite{SimMcCHag1999} and vice versa in crowded environments. These phase transitions are called solitarization and gregarization. Gregarization dominates when large numbers of locusts gather at the same site, potentially leading to a destructive outbreak \cite{DesColSim2000,ColDesSim1998}.

Locust gregarization may be induced by visual, olfactory, or tactile cues. For the desert locust \emph{Schistocerca gregaria}, the most potent stimulus is tactile: repetitive stroking of the femora of hind legs \cite{SimMcCHag1999,SimDesHag2001,RogMatDes2003} functions as a crowding indicator.  Mechanosensory stimulation of leg nerves leads to serotonin cascades in the metathoracic ganglion, and initiates gregarious behavior \cite{SimDesHag2001,RogMatDes2003,AnsRogOtt2009}. Gregarization can be induced by rubbing a locust's hind leg for $5\ s$ per minute during a period of $4\ hr$ \cite{SimDesHag2001}. Cessation of physical contact leads to solitarization after $4\ hr$, though the degree of solitarization achieved during that time depends on the individual's ancestry.

Experiments and models have shed much light on how group alignment \cite{RomCouSch2009,BuhSumCou2006,Sum2010,YatErbEsc2009} and group motion \cite{BazBuhHal2008,BazRomTho2011} depend on group size or density and treatments such as diet and denervation.  For instance, a low-protein diet (which motivates cannibalism in locusts)  leads to stronger interactions between individuals and lowers the threshold density beyond which mean speed and group coherence increase \cite{BazRomTho2011}. Other data-driven studies include models based on a well-known physics paradigm for self-propelled particles \cite{VicCziBen1995} and explore the transition between a disordered and a coherent marching group. Both \cite{EdeWatGru1998} and \cite{TopBerLog2008} study the dynamics of rolling patterns formed by flying, gregarious swarms. A logistic map was introduced in \cite{HolChe1996} to describe phase change via a birth rate and a carrying capacity dependent on population density modulated by stochastic effects. 

Our current work complements previous locust modeling studies in several ways. First, many of the previous models are individual-based (Lagrangian) simulations, where the position, velocity, and interactions of individual locusts are tracked\cite{RomCouSch2009,BuhSumCou2006,YatErbEsc2009,BazBuhHal2008,BazRomTho2011}. Ours is density-based (Eulerian), allowing  techniques of partial differential equations (PDEs) and their extensions (integro-PDEs) to be utilized. Second, we concentrate on gregarious-solitarious transitions not yet explicitly considered in \cite{BuhSumCou2006,BazRomTho2011}. We address intrinsic attractive-repulsive social interactions, whereas many current models consider interactions with clumped resources and environmental heterogeneity as their focal points \cite{ColDesSim1998,DesColSim2000}. Finally, some models \cite{BazRomTho2011} include anisotropic interactions such as different responses to anterior and posterior neighbors, or consider Newtonian dynamics. To explore minimal mechanisms sufficient for band formation, our work instead uses isotropic interactions and a kinematic approach. The open problem we address via mathematical modeling is to quantify and describe collective gregarization, a key, early process that necessarily occurs before the emergence of a destructive locust outbreak. We do this by linking the physiology of individual-level phase change and interphase interactions to predictions at the level of the gregarious hopper band as a whole.

We investigate the onset of an outbreak by constructing a continuum mathematical model of behavioral phase for interacting gregarious and solitarious locusts. We classify and quantify group dynamics in wide swaths of parameter space, a task which is challenging by numerical techniques alone. We find that in the limit of low densities, both phases are uniformly spread and the solitarious phase dominates. For sufficiently large populations, a dense, traveling patch of gregarious locusts suddenly emerges, while solitarious locusts become more and more scarce. We identify locust clustering at high densities with the onset of a hopper band. Through analysis of our model, we calculate the critical density beyond which the gregarious group forms, and for the final ratio of gregarious to solitarious locusts. We determine these quantities in terms of behavioral parameters at the level of individual locusts, hence connecting individual and group properties. Our model also displays population-level hysteresis, which has implications for locust management.

% Results and Discussion can be combined.

\section*{Model}

\subsection*{Model construction}

Locusts in a group are subject to attractive and/or repulsive forces based on combined sensory, chemical, and mechanical cues that affect their motion. We assume that sensing is directionally isotropic, a reasonable approximation \cite{ParMar2005} for organisms receiving sensory inputs of a variety of types, although directional models are possible as well \cite{EftVriLew2007}.  Rather than tracking individual locusts, we consider a population density field $\rho(\vec{x},t)$ moving at velocity $\vec v(\vec x,t)$.  Continuum population modeling
\cite{KelSeg1971b, Oku1980} allows us to apply analytical tools in order to characterize swarm formation and structure. Our work draws from classic swarm modeling in which a conserved population density field $\rho$ moves at a velocity $\vec v$ that arises from social interactions:
\label{eq:social}
\begin{equation}
\rho_t + \nabla \cdot (\rho \vec{v}) = 0. \label{eq:sociala}
\end{equation}
This is the well-known mass balance equation that tracks individuals moving collectively at velocity  $\vec{v}$. It is typically assumed that individuals can sense the population density nearby, and that this sensing gives rise to attractive-repulsive social forces $\vec{F}$, or alternatively, social potentials $Q$ (the negative gradients of which are forces). Within this context, the contribution $\rho(\vec{x}',t)$ of a small clump of individuals at location $\vec{x}'$ to the force on the individual at position $\vec{x}$ is given by $\vec{F}(\vec{x}-\vec{x}') \rho (\vec{x},'t) = -\nabla Q (\vec{x}-\vec{x}') \rho(\vec{x}',t)$. The corresponding velocity is proportional to the forces exerted by neighbors at all spatial locations, so that $\vec{v}(\vec{x},t)$ is given by integration over all $\vec{x}'$ as
\begin{equation}
\vec{v}(\vec{x},t) = -
\int_\Omega {\nabla Q}(\vec{x}- \vec{x}')  \rho (\vec{x}',t)\,
d\vec{x}'. \label{eq:socialb}
\end{equation}
The expression for the velocity $\vec v(\vec{x},t)$ in Eqn.~\eqref{eq:socialb} is a convolution of the density $\rho(\vec x,t)$ and the social interaction force $-\nabla Q(\vec x - \vec x')$, which describes the influence of the locust population at location $\vec{x'}$ on that at location $\vec{x}$. This is a common formulation of so-called \emph{nonlocal} interaction models \cite{MogEde1999,TopBerLew2006,LevTopBer2009,BerTop2011}, which capture interactions that are spatially distributed, in contrast to pure partial differential equations, which include only local terms such as derivatives and gradients, and which describe interactions only over infinitesimal ranges. Nonlocal aggregation models have been studied for various social interactions $Q$; known solutions include steady swarms, spreading populations, and contracting groups. We use the notation $\vec v = - \nabla Q * \rho$ to denote the convolution in Eqn.\,\eqref{eq:socialb}. We assume that $Q (\vec x - \vec x')$ is radially symmetric and depends only on the distance between $\vec x$ and $\vec x'$. The detailed forms of $Q$ in the case of solitarious and gregarious locusts will be described later.

To adapt Eqs.\,\eqref{eq:sociala} and \eqref{eq:socialb} to biphasic insects, we introduce separate density fields for solitarious and gregarious locusts, $s(\vec{x},t)$ and $g(\vec{x},t)$, respectively, and the total local density $\rho = s+g$. With marching locusts in mind, we consider a two-dimensional geometry, with $\Omega$ representing the spatial domain and $\vec{x}= (x,y)$ as spatial coordinates. We now include the phase transitions between solitarious and gregarious locusts. To do so, we define two density-dependent functions,  $f_1(\rho)$ for the  the rate of gregarious-to-solitarious transition, and $f_2(\rho)$ for the rate of solitarious-to-gregarious transition. Our model thus reads
\begin{subequations}
\label{eq:ge}
\begin{alignat}{4}
\dot{s} &+ \nabla \cdot (\vec{v}_s s)  &=  -f_2(\rho)s &+ f_1(\rho)g, \\
\label{eq:ge2}
\dot{g} &+ \nabla \cdot (\vec{v}_g g) &=  \phantom{-}f_2(\rho)s &-
f_1(\rho)g,  
\end{alignat}
\end{subequations}
where the velocities are given by
\begin{equation}
\label{eq:v}
\vec{v}_s = -\nabla(Q_s * \rho), \quad \vec{v}_g =  -\nabla(Q_g * \rho).
\end{equation}
These equations are complete once we specify the solitarious and gregarious social interactions $Q_{s,g}$ and the density-dependent conversion rates $f_{1,2}$. Since solitarious locusts are crowd-avoiding, we take $Q_s$ to be purely repulsive. Gregarious locusts, on the other hand, are attracted to others, except for short-distance repulsion due to excluded volume effects. Hence, we model $Q_s$ and $Q_g$ as
\begin{equation}
\label{eq:Q}
Q_s(\vec{x}-\vec{x}') = R_s \exp{-|\vec{x}-\vec{x}'|/r_s}, \quad Q_g(\vec{x}-\vec{x}') = R_g
\exp{-|\vec{x}-\vec{x}'|/r_g} - A_g \exp{-|\vec{x}-\vec{x}'|/a_g},
\end{equation}
where $R_s, R_g, A_g$ are interaction amplitudes that determine the strengths of attraction and repulsion, and $r_s, r_g$ and
$a_g$ are interaction length scales that represent typical distances over which one locust can sense and respond to another.

The above forms of $Q_{s,g}$ describe social interactions that decay exponentially away with distance from the sensing individual and are chosen to be isotropic for simplicity. As evident from Eqn.\,\eqref{eq:Q}, $Q_s$ is purely repulsive for all choices of $R_s$ and $r_s$. On the other hand, $Q_g$  is the difference of two exponentials, implying that there may be a distance at which repulsion and attraction balance, resulting in no net contribution to the velocity. The location of this balance point can be obtained by imposing $-\nabla Q_g = 0$ to obtain the critical distance
\begin{equation}
d = \frac{a_g r_g}{a_g - r_g} \ln \left(  \frac{R_g a_g}{A_g r_g} \right).
\end{equation}
Depending on the choice of social interaction parameters, the expression for $d$ may yield unphysical results such as negative distances. The distance $d$ also pertains only to two isolated locations $\vec{x}$ and $\vec{x}'$ and does not capture population-level features. Even for meaningful values of $d$, a collection of individuals interacting under $Q_g$ may disperse, aggregate, or clump. It is thus important to choose the appropriate parameter ranges for $a_g$, $r_g$, $A_g$ and $R_g$ so that the tendency of gregarious locusts to aggregate is modeled properly. Mathematical studies have shown that in order for cohesiveness to occur, the parameters in $Q_g$ must lie in a particular regime that leads to clumping \cite{DOrChuBer2006}. Thus, we require $R_g a_g - A_g r_g > 0$ so that repulsion dominates at short length scales, and $A_g a_g^2 - R_g r_g^2>0$ so that attraction dominates at longer ones. Taken together, these conditions guarantee a meaningful critical distance $d$ and macroscopic clumping behavior. We assume these conditions to hold for the remainder of this paper.

It remains to specify how density affects transitions from one phase to another. We call upon the biological observation that at higher densities, gregarization proceeds more quickly and solitarization more slowly. We model the phase conversion rates with the rational functions
\begin{equation}
\label{eq:rates}
f_1(\rho) = \frac{\delta_1}{1+ \left( \rho/k_1 \right)^2}, \quad
f_2(\rho) = \frac{\delta_2 \left( \rho/k_2 \right)^2}{1+ \left(
  \rho/k_2 \right)^2}.
\end{equation}
 The parameters $\delta_{1,2}$ are maximal phase transition rates and $k_{1,2}$ are characteristic locust densities at which $f_{1,2}$ take on half of their maximal values. Note that $f_1$ decreases with $\rho$, capturing the inverse relationship between solitarization rate
and density, while $f_2$ increases with $\rho$ and saturates at $\delta_2$, describing speedier gregarization at higher densities.

Our complete model consists of Eqs.\,\eqref{eq:ge}-\eqref{eq:rates} together with initial conditions specifying $s(\vec{x},0)$ and
$g(\vec{x},0)$. We consider a spatially periodic domain, which simplifies both numerical simulation and mathematical analysis. In certain laboratory studies using ring-shaped arenas, such boundaries are natural (while being less ideal for comparison with field studies) \cite{BuhSumCou2006}. We do not include locust reproduction or death as these occur on much longer time scales than phase change.

The model presented here is a general one containing some fundamental elements of locust dynamics. This work can be readily modified and extended to include details pertaining to different locust species, interactions with the surrounding environment, locust reproduction, and more. For instance, in our model, we have not explicitly accounted for the differing activity levels of solitarious and gregarious individuals \cite{PenSim2009}. Additionally, while gregarization is relatively fast for \emph{Schistocerca gregaria}, full solitarization may occur only after several generations of locusts. The phase conversions of \emph{Chortoicetes terminifera}, on the other hand, are characterized by similar timescales for the two phase conversions, so that both gregarization and solitarization occur rapidly within the lifetime of a single locust individual \cite{GraSwoAns2009}. On another note, vegetation or waterway patterns may impose spatial inhomogeneities such as non-uniform initial distributions of solitarious locusts, or attraction to preferred sites. Preexisting models in the literature have pointed out the important link between the spatial distribution of vegetation, as well as nutritional quality, on locust clustering, gregarization, and swarming \cite{ColDesSim1998,DesColSim2000,DesSim2000b,DesSim2000}. All of these elements could be used to refine our model for predictive purposes. However, as the first work in the continuum modeling of locust population phase change, ours begins with the fundamental model contained in Eqs.\,\eqref{eq:ge}-\eqref{eq:rates}. Our model is complementary to the preexisting ones in that we focus on how inherent inter-individual interactions can lead to gregarization and swarming, even in a spatially homogeneous environment. Multi-generational dynamics, differential activity levels, resource distribution, and related factors could be considered as possible extensions of our work.

\subsection*{Parameter selection}

Some of our results are analytical formulas, which may be evaluated for any desired parameter values. Other results depend on numerical computations, and these require specific choices of parameters. For these results, we consider two different sets of phase transition parameters. (1) Most of our numerical results have been obtained using our \emph{default set} of parameters, based on estimates from the biological literature. Specifically we take $\delta_{1,2} = \delta = 0.25\ hr^{-1}$, corresponding to a gregarization time scale of approximately $1/\delta = 4\ hr$ for desert locusts (for whom some -- but not total -- solitarization occurs on the same time scale) \cite{SimDesHag2001,PenSim2009}. We also take $k_{1,2}=k = 65\ locusts/m^2$, since for desert locusts, the critical density for the onset of collective motion is $50\ \mbox{-}\ 80\ locusts/m^2$ \cite{BazRomTho2011}. We will allow for some deviation from $\delta_1 = \delta_2$ and $k_1 = k_2$ via a parameter sensitivity analysis. (2) To examine situations with large differences in the rates of gregarization and solitarization, we consider an \emph{alternative set} of parameters with $\delta_1 = 0.025\ hr^{-1}$ and $\delta_2 = 0.25\ hr^{-1}$, so that gregarization is an order of magnitude faster that solitarization. We take $k_1 = 20\  locusts/m^2$ and $k_2 = 65\ locusts/m^2$ to model a gregarious-to-solitarious transition that occurs at a higher density threshold than the solitarious-to-gregarious transition.

We use the same social interaction parameters for all results (variations from this set are accounted for by a sensitivity analysis). To estimate the social interaction length scale parameters in Eqs.\,\eqref{eq:Q}, we apply the results of \cite{BuhSumCou2006,BazRomTho2011}, which identify the ``sensing range'' of a desert locust as $0.14\ m$, and the ``repulsion range'' as $0.04\ m$, close to the approximately $0.05\ m$ body length of a desert locust at the fifth instar of its development. For the gregarious phase we thus set the repulsion length scale at $r_g = 0.04\ m$ and the attractive one at $a_g = 0.14\ m$, corresponding to the experimental sensing range. These choices agree with theoretical studies showing that for cohesive swarms,  attraction occurs over longer length scales than repulsion \cite{MogEde1999,MogEdeBen2003}. We also assume that solitarious locusts are repelled from others at their sensing range, so that $r_s = 0.14\ m$. These choices satisfy $r_g < a_g = r_s$ which is assumed for the remainder of this paper.

Finally, we estimate $R_s$, $R_g$, and $A_g$ via explicit velocity computations. The speed of a locust when it is alone varies between $72\ \mbox{-}\ 216\ m/hr$, depending on diet \cite{BazRomTho2011}. At the upper end, this is roughly one body length per second. When it is moving in a group, the individual's speed varies in a tighter range of $144\ \mbox{-}\ 216\ m/hr$ \cite{BazRomTho2011}. In making our phase-dependent velocity estimates, we interpreted the ``moving alone'' and ``moving in a group'' data as typical to solitarious and gregarious locusts, respectively. Using these biological measurements and Eqn.\,\eqref{eq:v}, we find $R_s = 11.87\ m^3/(hr \cdot locust)$, $R_g = 5.13\ m^3/(hr\cdot locust)$, and $A_g = 13.33\ m^3/(hr\cdot locust)$. Details are given in Text S1. Our choices of social interaction parameters satisfy conditions mentioned in the previous section, namely $R_g a_g - A_g r_g > 0$, and $A_g a_g^2 - R_g r_g^2>0$ so that gregarious insects will clump.

Most of our parameter choices have been inferred or estimated from published laboratory experiments. It is possible however, that in the field, some parameter values may be quite different from the ones we have used. For instance, locusts in the field may pause while marching to perch on the vegetation, giving rise to an effective speed that is lower than what measured in lab experiments, where perching does not occur. It is also noteworthy that gregarious locusts are more active than solitarious locusts, a fact that is reflected by our method of choosing $R_s, R_g, A_g$  from estimates of the velocities of individuals when moving alone and in a group. As we describe below, we analyze our model varying all parameters within reasonable bounds: our results are qualitatively the same.

\section*{Results}

We first determine the simplest solutions to the model, namely those for which the densities of gregarious and solitarious locusts are in a spatially uniform steady state. We probe the stability of that uniform state using linear stability analysis (LSA), a calculation that addresses whether small, spatially nonuniform perturbations grow or decay. This is equivalent to determining the signs of eigenvalues of the linearized system, where positive (negative) eigenvalues imply growing (decaying) perturbations. The rate of initial growth/decay depends on the wavenumber of the perturbation. The growing perturbations can be interpreted in terms of nascent aggregates of locusts, and the wave numbers as the number of aggregates per unit area. The analysis provides a condition for the onset of aggregation, namely the emergence of positive eigenvalues of the linearized model. In our case, this aggregation condition is shown below in Eqn.\,\eqref{eq:cond2}. LSA cannot, in general, predict the ensuing dynamics once perturbations have grown to a large size. Further analysis uses an approximation to eliminate the spatial dependence of the model, which enables an analytical prediction of the proportion of solitarious and gregarious locusts on a longer time scale. To visualize the dynamics of aggregation, we perform numerical simulations in one spatial dimension using the linear stability analysis to identify regimes of interesting behavior. The model displays population-level hysteresis.

\subsection*{Homogeneous steady states}

The solitarious $s_0$ and gregarious $g_0$  homogeneous steady-state (HSS) solutions of Eqn.\,\eqref{eq:ge}  can be written in terms of the total uniform density $\rho_0$, which is simply the mean value of $\rho$ for a specified initial condition. The full expressions for $s_0$ and $g_0$ in terms of $\rho_0$ appear in Text S1; in the small $\rho_0$ limit these are approximately
\begin{equation}
s_0 \approx \rho_0 - \frac{\delta_2}{\delta_1 k_2^2} \rho_0^3, \quad
g_0 \approx \frac{\delta_2}{\delta_1 k_2^2} \rho_0^3,
\end{equation}
while in the limit of large $\rho_0$ we find 
\begin{equation}
s_0 \approx \frac{\delta_1 k_1^2}{\delta_2 \rho_0}, \quad g_0 \approx 
\rho_0 -\frac{\delta_1 k_1^2}{\delta_2 \rho_0}.
\end{equation}
The low density HSS is thus composed mostly of solitarious locusts and vice versa for the high-density case, showing the non-monotonicity of $s_0$ with respect to total density $\rho_0$. In Fig.\,\ref{fig:steadystate}(A) we plot the HSS $s_0$ (middle solid blue curve) and $g_0$ (middle broken green curve) for our default set of phase change parameters, $k = 65\ locusts/m^2$ and $\delta = 0.25\ hr^{-1}$.

As shown, $s_0$ initially increases with $\rho_0$. At a critical density $\rho_*$, $s_0$ reaches a maximum, whereas $g_0$ keeps increasing monotonically. Fig.\,\ref{fig:steadystate}(B) shows a blow-up of the region near $\rho_*$. For our default parameters, the maximum value $s_0^{\rm max}$ is attained at $\rho_* =k$, the same density value for which solitarious and gregarious densities coincide so that $s_0^{\rm max} = s_0(\rho_*) = g_0(\rho_*) = k/2$. However, this feature is a result of our choice $k_1=k_2$ and $\delta_1=\delta_2$. In general, the point of maximum solitarious density and the point of equal solitarious and gregarious density do not coincide, as is directly deducible from the full expressions for $s_0$ and $g_0$ in Text S1. To give a sense of detuning from our parameter estimates, we also calculate and plot $s_0$ and $g_0$ for parameter sets chosen randomly from uniform distributions
centered at our estimated default set of values for $\{ \delta_1, \delta_2, k_1, k_2\}$. The bottom and top curve in each set show the 25th and 75th percentile values.

We also study a much more general case where $ \delta_1 \ne \delta_2, k_1 \ne k_2 $, in keeping with the distinct rates of transition and critical transition densities seen biologically. As an alternative way to understand the HSS solutions, we consider the fractions $\phi_{s,g}$ of solitarious and gregarious locusts, where $\phi_s + \phi_g = 1$. As shown in Text S1, for the HSS,
\begin{equation}
\label{eq:phigrho}
\phi_g = \biggl\{1 + \gamma K^2 \frac{1+\psi^2}{\psi^2(\psi^2+K^2)}\biggr\}^{-1}.
\end{equation}
Here, $\gamma = \delta_1/\delta_2$ is the ratio of maximal solitarization rate to maximal gregarization rate,  $K = k_1/k_2$ is the ratio of the characteristic solitarization and gregarization densities for individuals, and $\psi = \rho_0/k_2$ is a rescaled spatially homogeneous density. The gregarious fraction $\phi_g$ is monotonically increasing in $\psi$, and hence in $\rho_0$; that is to say, as total density increases, the gregarious fraction increases. For small $\rho_0$, $\phi_g \approx 0$, but as $\rho_0$ increases, there is a crossover between solitarious and gregarious populations. Uniformly spread solitarious populations cannot be sustained when the density is too high: the gregarious state will necessarily become the dominant one.

\subsection*{Linear stability analysis}

To determine conditions under which a  nearly uniformly spread locust population aggregates or disperses, we study the linear stability of the HSS (details appear in Text S1). The calculation is a standard but somewhat tedious exercise. In nonlocal systems such as ours, linear stability results depend on the Fourier transforms $\wh{Q}_{s,g}(q)$ of the interaction potentials $Q_{s,g}$. For our locust model, the stability of the HSS depends on the eigenvalue
\begin{equation}
\label{eq:lambda1}
\lambda_1(q) = -q^2 \left[s_0 \wh{Q}_s(q) + 
g_0 \wh{Q}_g(q)\right],
\end{equation}
where $q = |\vec{q}|$ is the perturbation wave number and the Fourier transforms $\wh{Q}_{s,g}(q)$ in two dimensions are
\begin{eqnarray}
\label{eq:qhats}
\widehat{Q}_s(q)& =&  \frac{2 \pi R_s r_s^2}{(1+r_s^2 q^2)^{3/2}}, \\
\widehat{Q}_g(q) & = &   \frac{2 \pi R_g r_g^2}{(1+r_g^2q^2)^{3/2}}-
\frac{2 \pi A_g a_g^2}{(1+a_g^2 q^2)^{3/2}}.
\end{eqnarray}
Observe that the eigenvalue $\lambda_1(q)$ depends on all of the individual-based parameters governing rates of phase change (via $s_0$ and $g_0$) and all of the social interaction amplitudes and length sensing length scales. The HSS derived in the previous section is stable to small perturbations if $\lambda_1(q)<0$ for all $q$. If $\lambda_1(q) > 0$ for some $q$, then the HSS is unstable to perturbations of those wave numbers. 

\begin{figure}[!ht]
\begin{center}
\includegraphics{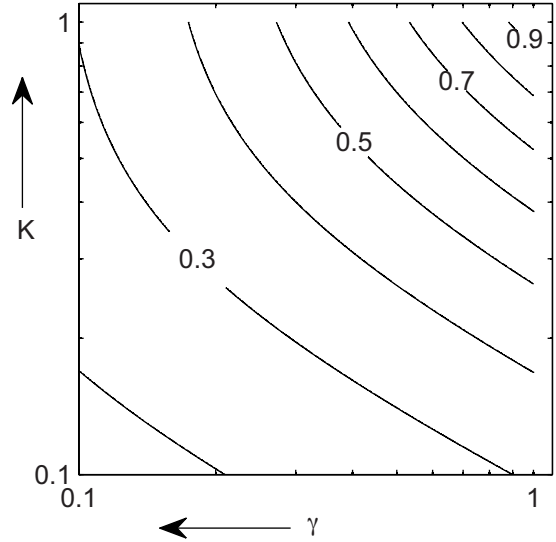}
\end{center}
\caption{{\bf Linear stability of spatially homogeneous steady state (HSS) solutions.} The dimensionless phase change parameter ratios $\gamma = \delta_1/\delta_2$ and $K = k_1/k_2$ vary along the horizontal and vertical (we have used log axes). The contours indicate the critical value of rescaled density $\psi^* = \rho_0^*/k_2$. For rescaled densities greater than that value, the HSS solution is unstable. The critical rescaled density is monotonically increasing in both $\gamma$ and $K$. The arrow along the horizontal axis indicates the direction $\gamma$ moves if the relative rate of gregarization is increased (faster gregarization). The arrow along the vertical axis indicates the direction $K$ moves if the relative density threshold for gregarization is decreased (easier gregarization). For an accurate biological interpretation, one must multiply $\psi^*$ by $k_2$ in order to obtain the unscaled critical density $\rho_0^*$.\label{fig:linstab} }
\end{figure}

Our full analysis of this eigenvalue appears in Text S1. We formulate the instability condition in terms of $\phi_g$,
\begin{equation}
\label{eq:cond2}
\phi_g > \phi_g^* = \frac{R_s r_s^2}{R_s r_s^2 - R_g r_g^2 + A_g a_g^2}.
\end{equation}
If this condition is satisfied, initially small perturbations from the uniform steady state will grow. This inequality is a key result, and implies that if a sufficiently large fraction of the population is gregarious, the HSS solution is unstable. To obtain a more explicit condition in terms of the density $\rho_0$, one must substitute $\phi_g^*$ into Eqn.\,\eqref{eq:phigrho}, which relates gregarious fraction to total (scaled) density. One may then calculate the critical density $\rho_0$ above which the HSS is unstable. Since $\phi_g$ and $\rho_0$ are monotonically related, we conclude that the HSS solution is unstable for sufficiently dense populations. The algebra is tedious, and relegated to Text S1.  Instead, we present a contour plot in Fig.\,\ref{fig:linstab} which succinctly illustrates the stability features of the HSS. The phase change parameter ratios $\gamma = \delta_1/\delta_2$ and $K = k_1/k_2$ vary along the horizontal and vertical axes and the contours indicate the critical value of rescaled density $\psi^* = \rho_0^*/k_2$. For scaled densities greater than $\psi^*$, the HSS solution is unstable. The critical scaled density is monotonically increasing in both $\gamma$ and $K$. (Note that for an accurate biological interpretation, one must multiply $\psi^*$ by $k_2$ in order to obtain the unscaled critical density $\rho_0^*$.)

\begin{figure}[!ht]
\begin{center}
\includegraphics{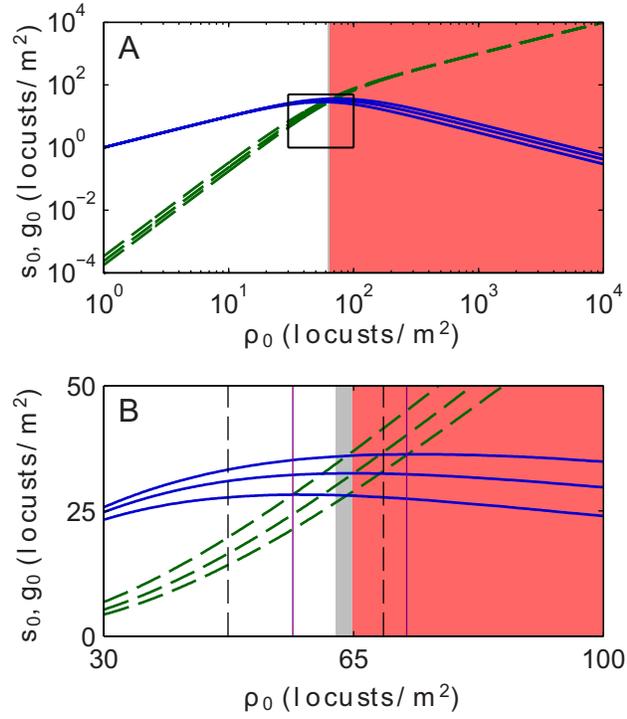}
\end{center}
\caption{{\bf Spatially homogeneous steady states (HSS).} (A) Spatially homogeneous solitarious ($s_0$, solid blue) and gregarious ($g_0$, broken green) steady state locust density as the total locust density ($\rho_0$) is varied. For each set of curves, the middle curve represents the solution for our default phase change parameters, $k_1=k_2= 65\ locusts/m^2$ and $\delta_1 = \delta_2 = 0.25\ hr^{-1}$. The bottom and top curve in each set show parameter sensitivity; they are the 25th and 75th percentile values for $s_0$ and $g_0$ over 10,000 parameter sets of $\{ \delta_1, \delta_2, k_1, k_2\}$ sampled from uniform distributions centered at the default values and varying by $\pm30$\%. In both the thin grey and red regions, the HSS is linearly unstable to small perturbations. Additionally, in the red region, $g_0 > s_0$, while in the grey and white regions, the opposite holds. (B) A blow-up of the boxed transition region in (A) around which the value of $g_0$ overtakes $s_0$. The dashed black vertical lines indicate the 25th and 75th percentile for this transition. The solid purple vertical lines indicate the 25th and 75th percentile values for the onset of linear instability.\label{fig:steadystate}}
\end{figure}

Upon inserting our default parameters in Eqn.\,\eqref{eq:cond2} we find that the homogeneous solution is unstable for $\rho_0 > \rho_0^* = 62.3\ locusts/m^2$. This value corresponds to the left border of the grey region in Fig.\,\ref{fig:steadystate}. For $\rho_0 > \rho_0^*$, to the right of the border, we expect the onset of a locust hopper band, \emph{i.e.},~formation of patches of high locust density that
can seed the clustering and gregarization of other locusts. In Fig.\,\ref{fig:steadystate}, linear instability can occur even at densities $\rho_0$ for which $s_0$ exceeds $g_0$ for our chosen parameters (represented by the center solid blue and center broken green curves). This result implies that the onset of instability leading to mass gregarization can take place even if solitarious locusts initially outnumber gregarious ones. We will later discuss mass gregarization in more detail. To visualize detuning from this set of parameters, we include the 25th and 75th percentile values of $\rho_0^*$ for onset of instability as vertical purple lines; these are again calculated by drawing 10,000 random samples of the parameters $k_{1,2}$, $\delta_{1,2}$, $R_{s,g}$, $r_{s,g}$, $A_g$, and $a_g$.  As seen from Fig.\,\ref{fig:steadystate}(b) our conclusions are robust across the randomly chosen parameter sets.

For our default set of biological parameters, $\phi_g^* \approx 0.479$ via Eqn.\,\eqref{eq:cond2} and $\rho_0^*$ turns out to be near $k=k_{1,2}$. We stress that generically, it is not the case that $\rho_0^*$ needs to be near $k_1$ and/or $k_2$. For our default  parameter set, $K = \gamma = 1$, in which case $\psi = 0.959$, so that the critical value $\rho_0^*$ is 95.9\% of $k_2$, namely $62.3\ locusts/m^2$. However, for different choices of $K$ and $\gamma$, drastically different outcomes are possible. For instance, for our alternative parameter set where $K \approx 1/3$ and $\gamma = 1/10$, the critical density is $\rho_0^* = 15.9\  locusts/m^2$, which is quite disparate from the individual gregarization density of $65\ locusts/m^2$, and is also less than the solitarization density of $20\ locusts/m^2$. Furthermore, for different choices of the social interaction parameters entering into Eqn.\,\eqref{eq:cond2}, it is possible to obtain a critical gregarious fraction $\phi_g^*$ that is much less than 1/2, meaning that instability and clumping can occur even with just a few gregarious insects.

\begin{figure}[!ht]
\begin{center}
\includegraphics{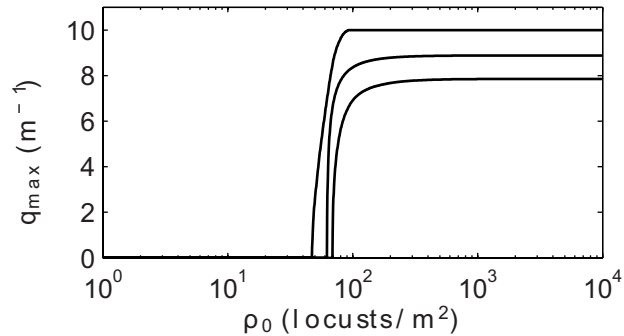}
\end{center}
\caption{{\bf Maximally unstable perturbation wave number $q_{\rm max}$ for homogeneous steady states with total density $\rho_0$.} Similar to Fig.~\ref{fig:steadystate}, the middle, bottom, and top curves show results for the 25th and 75th percentile as computed from 10,000 random parameter draws centered around our default parameter set. At low densities, there are no unstable perturbation wave numbers. Just past the critical density $\rho_0^*$, $q_{\rm max}$ increases rapidly and then plateaus. For our default parameters, $q_{\rm max}$ asymptotes to $8.89\ m^{-1}$ corresponding to a length scale of $0.71\ m$.\label{fig:stability}}
\end{figure}

For $\rho_0> \rho_0^*$, we can also find the wave number $q_{\rm max}$ corresponding to the most rapidly growing perturbation. Fig.~\ref{fig:stability} shows $q_{\rm max}$ for our chosen parameters (center curve) as well as the 25th and 75th percentile values over the 10,000 random parameter draws. The most unstable wave number $q_{\rm max}$ grows rapidly as a function of $\rho_0$ and then saturates at $q_{\rm max} \approx 8.89\ m^{-1}$, corresponding to a length scale $2 \pi / q_{\rm max} \approx 0.71\ m$ and indicating that the most quickly growing perturbations occur on the length scale of a few locust bodies.

Our linear stability analysis describes the behavior of small perturbations of uniform steady states, and is not expected to predict long-term or large-amplitude dynamics. For large perturbations, linear analysis is void. Additionally, even to analyze small perturbations of states other than uniform steady states, a different analysis would be needed.

\subsection*{Numerical simulation}

To illustrate the swarm dynamics described by Eqn.\,\eqref{eq:ge}, we simulate the model on a one-dimensional periodic domain of length $L=3\ m$ for a total population of $M=50$ locusts. Periodicity of the domain is an important aspect of a robust numerical platform devised for these simulations: we exploit the fact that convolutions $Q*\rho$  are easy to compute in Fourier space (where they are simply products, \emph{i.e.}, ${\hat Q} \cdot {\hat\rho}$), which significantly reduces the computational overhead. Computational issues associated with such convolutions also restrict us to one-dimensional simulations at present. At $t=0$ all locusts are solitarious and are randomly perturbed from the uniform density $s = M/L$, where $M$ is the total population mass
\begin{equation}
M = \int_\Omega \rho\,d\vec{x}.
\end{equation}
We adjust some parameters so as to adapt our model to the one-dimensional case. Specifically, one must take square roots of $k_{1,2}$ in order to collapse densities in a square to densities along a line segment. Consequently, for our default parameter set we choose $k_{1,2}=k=8\ locusts/m$ and $\delta_{1,2}=\delta = 0.25\ hr^{-1}$, whereas for the alternative set we use $k_1 = 4.5\ locusts/m$, $k_2 = 8\ locusts/m$, $\delta_1 = 0.025\ hr^{-1}$ and $\delta_2 = 0.25\ hr^{-1}$. In both cases we take the interaction amplitudes $R_s = 6.83\ m^2/(hr \cdot locust)$, $R_g = 6.04\ m^2/(hr \cdot locust)$, and $A_g = 12.9\ m^2/(hr \cdot locust)$, which have also been adapted from their original values to the one-dimensional case. The interaction length scales $r_s$, $r_g$, and $a_g$ are the same as for the two-dimensional case. Details of the numerical method and the parameter choices appear in Text S1.

\begin{figure}[!ht]
\begin{center}
\includegraphics{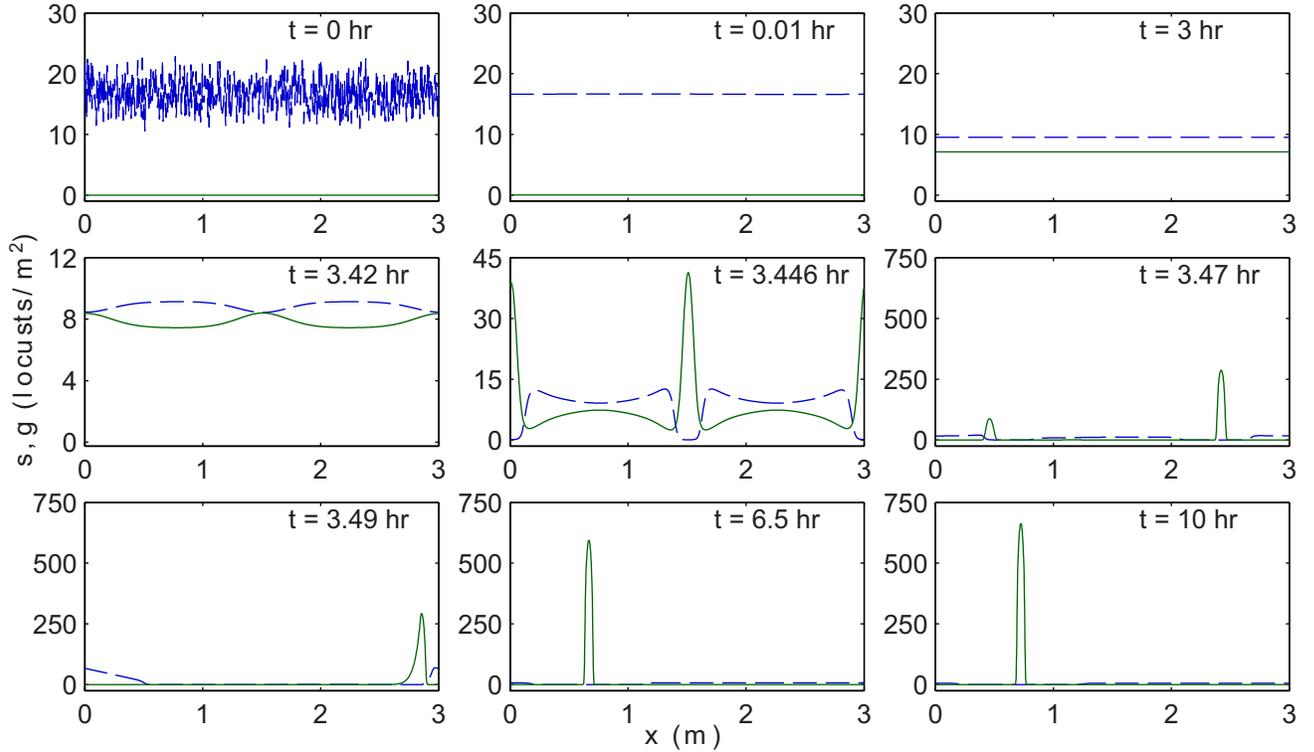}
\end{center}
\caption{{\bf Numerical simulations of Eqs.\,\eqref{eq:ge}.} Snapshots depict the numerical solution of Eqs.\,\eqref{eq:ge}-\eqref{eq:rates} at different times $t$ (in $hr$) on a periodic domain of length $L=3\ m$ with the default set of phase change parameters. See also Fig.\,\ref{fig:snapshotsNEW} for a comparison with the alternative parameter set. The solitarious (gregarious) density (in $locusts/m$) as a function of spatial position (in $m$) is shown in blue (green). The total population mass is $M = 50$ locusts and the initial condition is set at $g(x,t=0)=0$ and $s(x,t=0)$ given by a random perturbation centered around $s = M/L$. The top row of panels shows the fast smoothing of the initial state, and the subsequent evolution. Gregarization (approximately) occurs according to the spatially homogeneous version of Eqn.\,\eqref{eq:ge}, as can be seen up until the second row of panels, where the small instability becomes significant.  Two compactly supported clumps of gregarious locusts form, superposed on a very sparse population of solitarious insects. In the third row, the gregarious group travels as a propagating pulse, and eventually stops. During this stage, the gregarious and solitarious populations are essentially non-overlapping in space. As shown in Fig.\,\ref{fig:massfrac}, the group continues to slowly gregarize after it becomes stationary.\label{fig:snapshots}}
\end{figure}

\begin{figure}[!ht]
\begin{center}
\includegraphics{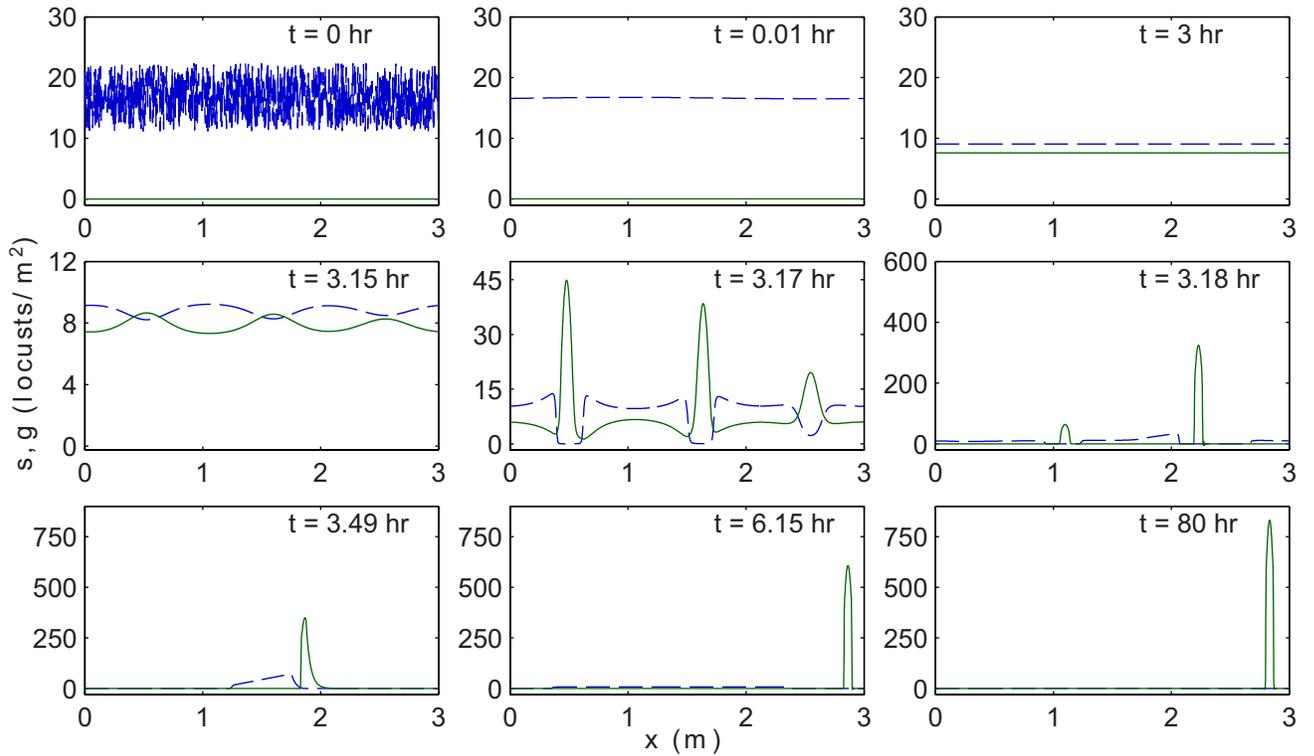}
\end{center}
\caption{{\bf Numerical simulations of Eqs.\,\eqref{eq:ge}.} Similar to Fig.\,\ref{fig:snapshots}, snapshots at different times $t$ (in hours), but for the alternative set of phase change parameters. Note that three, rather than two clumps of gregarious locusts form at intermediate times. This simulation is continued until $t=80\ hr$ (last frame) to show the stability of the final cluster of gregarious locusts.\label{fig:snapshotsNEW}}
\end{figure}

Results are shown in Fig.\,\ref{fig:snapshots} for the default parameter set and in Fig.\,\ref{fig:snapshotsNEW} for the alternative set. In each case, the snapshots show $s(x,t)$ (dashed blue curve) and $g(x,t)$ (solid green curve) at selected times. Starting from the randomized solitarious state at $t=0\ hr$, locusts rapidly redistribute to a roughly spatially uniform density until $t \approx 3\ hr$. Tiny variations are present but not visible on the scales of these figures. Gregarization and subsequent rapid spatial segregation follow. In Fig.\,\ref{fig:snapshots}, between $t \approx 3.42\ hr$ and $t \approx 3.47\ hr$, two compactly supported clumps of gregarious locusts emerge, superposed on a background of sparse, solitarious individuals.  A similar transition occurs between $t \approx 3.15\ hr$ and $t \approx 3.17\ hr$ in Fig.\,\ref{fig:snapshotsNEW}, but for these parameter values, we find initial clustering with three, rather than two density peaks. The number (or alternatively, length scale) of transient clumps that form appears to be selected dynamically. This intermediate dynamical selection process and the coarsening that ensues are avenues for future numerical and analytical investigation. In each example, the disjoint clusters quickly merge due to the long-range attraction of gregarious individuals. A single remaining pulse is formed by $t\approx 3.49\ hr$ in both cases and travels until $t \approx 6.5\ hr$, at which time the majority, but not all, of the solitarious locusts have transitioned to the gregarious form. Gregarization continues during the subsequent hours, albeit at a slower rate. For both figures, the gregarization of the final clump continues slowly, approaching an equilibrium at exponentially long times.

To study the locust gregarization process further, we define  the total mass of solitarious and gregarious locusts, $S$ and $G$, as
\begin{equation}
S =  \int_{\Omega} s\,d\vec{x}, \quad G =  \int_{\Omega} g\,d\vec{x},
\label{WasEq29}
\end{equation}
so that the total population mass is $M = S+G$. We also define the mass fractions
\begin{equation}
\label{eq:whsg}
\phi_s = S/M, \quad \phi_g = G/M, \quad \phi_s + \phi_g = 1,
\end{equation}
which we before calculated for HSS solutions, but we now generalize for spatially varying states. These quantities will be useful to further our mathematical analysis. Fig.\,\ref{fig:massfrac} shows $\phi_s(t)$ (blue curve) and $\phi_g(t)$ (green curve) as arising from the numerical simulations depicted in Fig.\,\ref{fig:snapshots} and  Fig.\,\ref{fig:snapshotsNEW}. Several distinct regimes are visible, and we discuss these below.

\subsection*{Spatially-homogeneous and spatially-segregated bulk theories}

\begin{figure}[!ht]
\begin{center}
\includegraphics{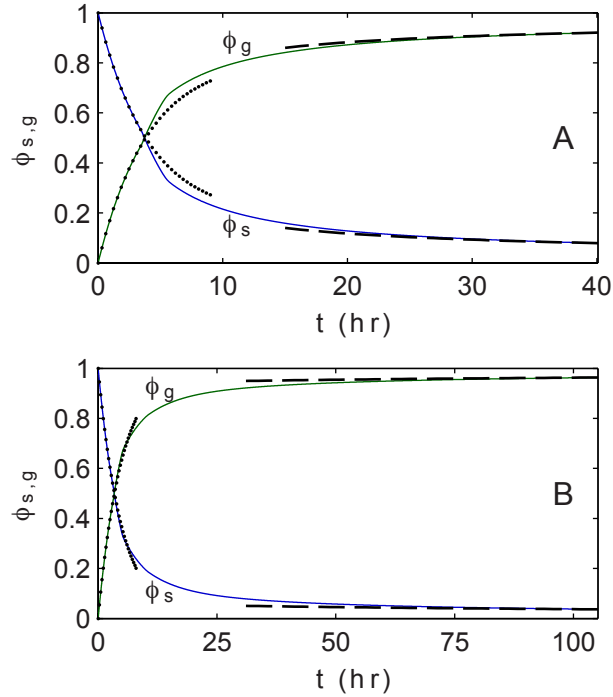}
\end{center}
\caption{{\bf Population-level phase change over time.} Mass fractions $\phi_{s},\phi_{g}$ of solitarious (blue) and gregarious (green) locusts as a function of time (in hours) for the numerical simulations of Fig.\,\ref{fig:snapshots} and Fig.\,\ref{fig:snapshotsNEW}. (A) Default set of phase change parameters, corresponding to the simulation in Fig.\,\ref{fig:snapshots}. (B) Alternative set of phase change parameters, corresponding to the simulation in Fig.\,\ref{fig:snapshotsNEW}. For both cases, at early times, these mass dynamics are well-approximated by the spatially homogeneous version of the governing equations Eqs.\,\eqref{eq:ge}, whose solution, Eqn.\,\eqref{eq:homogeneousdynamics}, is shown as dotted curves. At late times, the mass dynamics are approximately described by the spatially segregated bulk theory of Eqn.~\eqref{BulkModelEqs}, whose solution is shown as dashed curves.\label{fig:massfrac}}
\end{figure}

As visible in the second and third panels of Fig.\,\ref{fig:snapshots} and Fig.\,\ref{fig:snapshotsNEW}, the early-time dynamics of Eqs.\ \eqref{eq:ge} are approximately spatially homogeneous. As a result, spatially-dependent terms in Eqs.\ \eqref{eq:ge} are negligible, $\rho$ is approximately constant, and hence the governing equations are linear ordinary differential equations (ODEs) that are easily solved. We write the solution of these ODEs in terms of the mass fractions $\phi_{s,g}$,
\begin{equation}
\label{eq:homogeneousdynamics}
\phi_g(t) = \frac{f_2(\rho_0)}{f_1(\rho_0)+f_2(\rho_0)} \left\{1 - \exp{-[f_1(\rho_0)+f_2(\rho_0)]t}\right\}, \quad \phi_s(t) = 1 - \phi_g(t),
\end{equation}
where we have used the initial condition $\phi_s(t=0) = 1$. This analytical solution is plotted in Fig.\,\ref{fig:massfrac} as a dotted line, and agrees closely with the numerical results for the first few hours.

In the later panels of Fig.\,\ref{fig:snapshots} and Fig.\,\ref{fig:snapshotsNEW}, gregarious and solitarious locusts spatially segregate into areas with disjoint support. This means that in each distinct region, $\rho(x,t) \approx s(x, t)$ or $\rho(x,t) \approx g(x, t)$. We thus consider a \emph{bulk} model reduction to study the dynamics of the two non-overlapping solitarious and gregarious populations. In particular, we assume that solitarious locusts are spread throughout most of the domain $\Omega$, covering an area denoted $\alpha_s$, whereas gregarious locusts are confined to a region with area $\alpha_g$.  Within these areas, local densities are approximately $S/\alpha_s$ and $g = G/\alpha_g$. By integrating Eqs.\ \eqref{eq:ge} over the domain and assuming that $s$ and $g$ are approximately constant in their support, we obtain
\begin{equation}
\label{BulkModelEqs}
\frac{d\phi_s}{dt} = - \frac{c_1\phi_s^3}{1+c_2\phi_s^2} 
+ \frac{c_3 \phi_g}{1+c_4 \phi_g^2} = -
\frac{d\phi_g}{dt},
\end{equation}
where
\begin{equation}
c_1 = \frac{\delta_2 M^2}{\alpha_s^2 k_2^2} , \quad c_2 =
\frac{M^2}{\alpha_s^2 k_2^2}, \quad c_3 = \delta_1, \quad c_4 =
\frac{M^2}{\alpha_g^2 k_1^2}.
\end{equation}
The numerical solution of these ODEs (dashed lines in Fig.\,\ref{fig:massfrac}) agrees closely with the late time full-scale numerical simulation results, where we use values of $\alpha_{s,g}$ measured empirically from the terminal equilibrium.  One can reduce Eqn.\,\eqref{BulkModelEqs} to a single nonlinear ODE using $\phi_s=1-\phi_g$, though this equation is not amenable to analytical solution. Since we are interested in the large population limit for which we expect potential large scale gregarization, we instead study Eqs.\,\eqref{BulkModelEqs} for large $M$. In this case, to leading order in $M$, the bulk model reduces to
\begin{equation}
\label{BulkModelLargeM}
 \dot{\phi_s} =  - \delta_2 \phi_s +  \frac{c_3 }{c_4 \phi_g} = 
- \dot{\phi_g}. 
\end{equation}
Given the expressions for $c_{3,4}$ and the fact that $\phi_s, \phi_g \le 1$, the first term is $O(1)$ whereas the second one is much
smaller, $O(1/M^2)$. For large $M$ then, and to leading order, $\phi_s$ decays exponentially in time with rate $\delta_2$. This result is based on the assumption of a segregated state, and thus would be expected to occur only once segregation is nearly complete.

Since for large $M$ (nearly) the entire population will eventually become gregarious, the critical density $\rho_0^*$ is a crucial result. If the population is in the stable regime (where $\rho_0 < \rho_0^*$) then mass gregarization can be avoided and solitarious and gregarious locusts can coexist as uniformly spread populations.  However, as soon as the population shifts beyond the border of stability (where $\rho_0 > \rho_0^*$) the group gregarizes and the onset of a locust hopper band is inevitable.

\subsection*{Phase change and hysteresis}

The biological literature discusses the importance of \emph{hysteresis} in locust phase change, as reviewed, for instance, in \cite{PenSim2009}. It is important to disambiguate the possible meanings and interpretations of phase change hysteresis, to place this phenomenon within the context of our model, and most especially, to distinguish between hysteretic features at the individual and population levels.

One type of hysteresis is simply defined as ``rates of gregarization [that] differ from rates of solitarization''  \cite{PenSim2009}. Within our model, this type of hysteresis may be interpreted as cases where $\delta_1 \neq \delta_2$ or $k_1 \neq k_2$. Our results thus far have accounted for this type of hysteresis in three ways. First, for our primary parameter set in which $\delta_1 = \delta_2$ and $k_1 = k_2$, we have allowed deviations from equality by performing a sensitivity analysis incorporating variations of up to 30\% from the base parameter values, as represented in the results of Fig.\,\ref{fig:steadystate} and Fig.\,\ref{fig:stability}. Second, for our alternative parameter set, we have chosen $\delta_2 = 10 \delta_1$ and $k_2 \approx 3 k_1$. And finally, for analytical results such as the homogeneous steady states and their stability, we have obtained analytical formulas into which any values of $\delta_{1,2}$ and $k_{1,2}$ can be substituted.

Another interpretation of hysteresis relates to ``solitarization [having] two phases: an initial rapid phase and a second, slower phase that requires insects to be maintained in isolation across successive moults -- or generations'' \cite{PenSim2009}. Our model is constructed on the time-scale of a single generation, and thus we cannot account for this type of hysteresis, which would require a multi-generational model.

\begin{figure}[ht!]
\begin{center}
\includegraphics{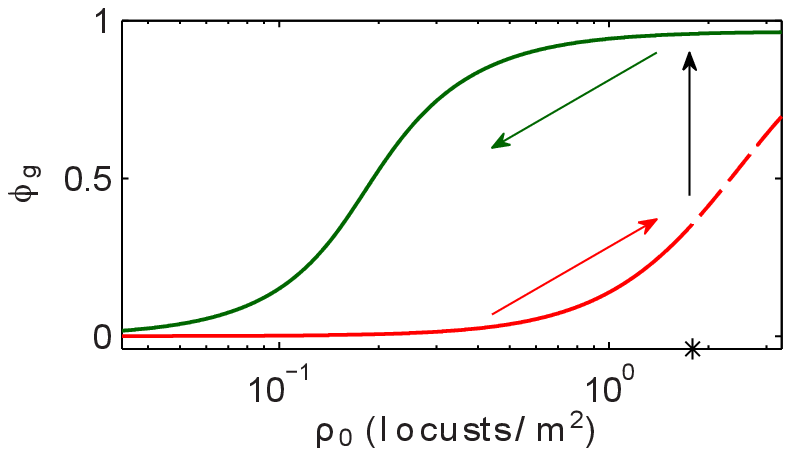}
\end{center}
\caption{{\bf Population-level hysteresis as a function of average density $\rho_0$.} Gregarious mass fraction $\phi_g$ as the average density $\rho_0$ (total mass $M$ divided by domain length $L$) is varied as a control parameter. We use our alternative set of phase change and social interaction parameters, as in Fig.\,\ref{fig:snapshotsNEW}. The solid (dotted) red curve represents the stable (unstable) homogeneous steady state solution, as calculated via linear stability analysis. As $\rho_0$ passes through the point of linear instability (marked with an asterisk) the solution jumps up to the green curve, which represents compactly supported gregarious aggregations obtained via numerical simulation, similar to the final states of Fig.\,\ref{fig:snapshots} and Fig.\,\ref{fig:snapshotsNEW}. As $\rho_0$ is decreased by slowly subtracting mass from aggregations on the green curve, the system remains on the upper branch even for values of $\rho_0$ sufficiently small as to be in the regime where the uniform state is stable, thus demonstrating dynamical population-level hysteresis.\label{fig:hysteresis}}
\end{figure}

Finally, we can consider population-level hysteresis. In the context of our model, this type of hysteresis refers to macroscopic properties of solutions of Eqn.\,\eqref{eq:ge} (which are outputs of the model) as opposed to differences in individual-level parameters (which are inputs to the model) as in the first type of hysteresis described above. Numerical results suggest that our model has population-level hysteresis; see Fig.\,\ref{fig:hysteresis}. This figure shows the gregarious mass fraction $\phi_g$ as the average density $\rho_0$ (total mass $M$ divided by domain length $L$) is varied as a control parameter. All phase change, social interaction, and physical domain parameters are the same as in Fig.\,\ref{fig:snapshotsNEW}.

The solid (dashed) red curve is an analytical result, representing the stable (unstable) HSS solution, as calculated previously via linear stability analysis. For small values of $\rho_0$, the HSS is stable to small perturbations. If locusts join the initially stable population the average density $\rho_0$ will increase (assuming a fixed spatial domain), shifting the uniform state to the right along the red curve; as yet no clustering will be evident. Beyond the point labeled with an asterisk, the uniform HSS loses stability and clustering occurs, as previously described. This corresponds to a jump represented by the vertical black arrow. The clustered state (green) is now stable. We next ask what happens if locusts are now removed from the aggregate, which corresponds to a reduction in $\rho_0$ (moving to the left in Fig.\,\ref{fig:hysteresis}). We answer this question numerically, by gradually subtracting mass from the population, allowing the system dynamics to evolve, and plotting the gregarious fraction as a function of mass. As the mass is slowly removed, the solution tracks leftwards along the green curve, indicating the persistence of the gregarious band. In fact, the band persists even partway into the regime where the HSS is linearly stable.

This dynamically observed hysteresis suggest that (for our model) a gregarious aggregation cannot be eliminated by reducing overall density to a low enough level where the HSS is linearly stable. This result has implications for locust control, as we discuss below.

\section*{Discussion}

In this paper,  we derived, analyzed, and simulated a model for the movement, social interactions, and density-dependent interconversions of the solitarious and gregarious forms of phase polyphenic locusts. The model is based on experimental observations and measurements, parameter values inferred from preexisting work, and basic assumptions about individuals' rules of behavior. We included social exchanges via repulsive and/or attractive interactions for gregarious and solitarious individuals, and we accounted
for phase change with density-dependent transitions, with crowding favoring solitarious-to-gregarious conversions. Our model was formulated in terms of continuum equations, allowing us to apply classical techniques such as linear stability analysis and bulk approximation. Since these methods were applied in two spatial dimensions, our results are relevant to insects aggregating in two dimensional structures such as hopper bands. We also provided example simulations in one spatial dimension as proof of principle, and as an indication of typical dynamics.

Our model explicitly takes into account intrinsic social interactions between individuals, in contrast to pre-existing models that focus on how insects respond to quality and spatial heterogeneity of nutrition or other environmental factors \cite{ColDesSim1998,BazRomTho2011,DesColSim2000}. These approaches are complementary, showing that both intrinsic and extrinsic factors that affect local densities also affect the gregarization transition.

Many of our results are achieved via mathematical analysis. The power of mathematical analysis is that it creates an explicit connection between individual-level and group-level quantities, \emph{e.g.}, via the inequality Eqn.\,\eqref{eq:cond2}. Once we identify the sensing range and interaction strength parameters in Eqn.\,\eqref{eq:Q} which govern individual locust attraction to and repulsion from others, we are able to calculate the critical density beyond which mass gregarization occurs.

Briefly, our results and predictions can be summarized as follows: (1) Locusts exist in a spatially uniform steady state distribution only up to a critical total population density. (2) Beyond this critical density, the uniform distribution can not be maintained, and massively dense gregarious clusters form. (3)~Linear stability analysis allows us to understand how the critical density depends on dimensionless ratios of the biological parameters. This dependence is summarized in Fig.\,\ref{fig:linstab}. Our analysis also yields the most unstable cluster spacing (from the wave number of the most unstable modes). (4) Numerical simulations illustrate the rapid transitions that take place once gregarization is initiated. Dense packs of gregarious locusts form and grow, and these move and sweep up solitarious locusts in their vicinity. (5) Via bulk approximation, we find estimates for the long-time mass fraction dynamics of solitarious and gregarious locusts. In the large population limit, the entire population will become gregarious. Bulk theory and simulations agree well, as shown in Fig.\,\ref{fig:massfrac}. (6) Our model displays population-level hysteresis, via which the critical density at which a gregarious aggregation forms from a dispersed population can be significantly higher than the density at which a gregarious aggregation would break up, as shown in Fig.\,\ref{fig:hysteresis}.

Our results shed light on locust control strategies in two ways. First, given the mass gregarization that takes place past the point of linear instability, the density threshold for this instability is a crucial quantity. In accordance with the idea proposed in \cite{SwoLecSim2010}, our work identifies a threshold below which populations should be kept in order to avoid a gregarious outbreak (assuming biological parameters are known to a sufficiently accurate degree). Furthermore, we have shown how this population-level property depends on individual-level parameters, finding a nontrivial relationship. Second, the apparent population-level hysteresis shows that dispersing a gregarized band, perhaps by killing individuals with pesticides, is harder than preventing group formation in the first place in that band annihilation requires a significantly lower locust density. In short, hysteresis implies that prevention could be more easily achieved than control.

Like all models, ours has its limitations. We did not include features of the environment such as vegetation, shown to have important influence on local crowding and hence gregarization.  Our simplifications lead to mathematical tractability, while limiting the direct biological relevance of the model at present. In the field, locusts encounter patchy vegetation and other environmental influences, and adding such factors to the model would make it more relevant to field experiments. Since we have not explicitly included resource gradients or other environmental cues, we do not here recapitulate the long-range motion of locust bands, but merely their formation and clustering. Including environmental factors constitutes an extension of the current framework. Similarly, simulations in two spatial dimensions are more challenging and remain open for future investigation.

Our work suggests several future biological experiments. First, as always, more accurate knowledge of model inputs would lead to better results. For our model, key inputs include the social interaction parameters, namely the length scales ($r_s$, $r_g$, and $a_g$) and interaction amplitudes ($R_s$, $R_g$, and $A_g$) in Eqn.\,\eqref{eq:Q} that we inferred from careful experiments such as those in \cite{BazRomTho2011}. However, to our knowledge, most of these parameters have not been directly measured in experiments on individuals. Second, we encourage observations of macroscopic group properties that could be compared to outputs of our model. These outputs include densities and sizes of bands. Additional quantitative field measurements along the lines of \cite{BuhSwoCli2011} could help validate and refine our model. Finally, we can imagine experiments that would probe important aspects of the system dynamics (as opposed to physical properties of the bands themselves). Hopper bands are known to undergo complicated dynamics, including splitting and merging \cite{Uva1977}. BBC video shows an example of such phenomena in \emph{Locustana pardalina} bands \cite{Locustvideo}. More accurate data for the dynamics of wild groups, including times for group formation and distances between merging bands and tributaries, could be compared to clumping time and length scales identified by our model. We are especially curious about experiments in which the critical average density for population-level gregarization and clumping might be probed in a controlled lab experiment, perhaps by slowly adding solitarious individuals into a large arena. Experimental measurements like those we have mentioned here would also motivate future two-dimensional extensions of our model where the streaming dynamics of hopper bands, the effects of the environment, and other stimuli could be more fully explored.

% You may title this section "Methods" or "Models". 
% "Models" is not a valid title for PLoS ONE authors. However, PLoS ONE
% authors may use "Analysis" 
%\section*{Materials and Methods}

% Do NOT remove this, even if you are not including acknowledgments
\section*{Acknowledgments}
This work was supported by a Wallace Scholarly Activities Grant from Macalester College (to CMT), National Science Foundation Grants DMS-1009633 (to CMT), DMS-1021850 and DMS-0719462 (to MRD), and an NSERC Discovery grant (to LEK). We are grateful to the Mathematical Biosciences Institute where portions of this research were performed. We thank the anonymous referees of this paper, whose feedback helped us tremendously.

%\section*{References}
% The bibtex filename
\bibliography{locustbib}

\begin{thebibliography}{10}
\providecommand{\url}[1]{\texttt{#1}}
\providecommand{\urlprefix}{URL }
\expandafter\ifx\csname urlstyle\endcsname\relax
  \providecommand{\doi}[1]{doi:\discretionary{}{}{}#1}\else
  \providecommand{\doi}{doi:\discretionary{}{}{}\begingroup
  \urlstyle{rm}\Url}\fi
\providecommand{\bibAnnoteFile}[1]{%
  \IfFileExists{#1}{\begin{quotation}\noindent\textsc{Key:} #1\\
  \textsc{Annotation:}\ \input{#1}\end{quotation}}{}}
\providecommand{\bibAnnote}[2]{%
  \begin{quotation}\noindent\textsc{Key:} #1\\
  \textsc{Annotation:}\ #2\end{quotation}}
\providecommand{\eprint}[2][]{\url{#2}}

\bibitem{Ken1951}
Kennedy JS (1951) The migration of the desert locust ({S}chistocerca gregaria
  {F}orsk.).
\newblock Proc Roy Soc Lond B 235: 163-290.
\bibAnnoteFile{Ken1951}

\bibitem{Alb1967}
Albrecht FO (1967) Polymorphisme Phasaire et Biologie des Acridiens Migrateurs.
\newblock Les Grands Probl{\`{e}}mes de la Biologie. Paris: Masson.
\bibAnnoteFile{Alb1967}

\bibitem{Uva1977}
Uvarov B (1977) Grasshoppers and Locusts, volume~2.
\newblock London, UK: Cambridge University Press.
\bibAnnoteFile{Uva1977}

\bibitem{Rai1989}
Rainey RC (1989) Migration and Meteorology: Flight Behavior and the Atmospheric
  Environment of Locusts and other Migrant Pests.
\newblock Oxford Science Publications. Oxford: Clarendon Press.
\bibAnnoteFile{Rai1989}

\bibitem{Bel2005}
Bell M (2005) The 2004 desert locust outbreak.
\newblock Bull Am Meteor Soc 86: S60.
\bibAnnoteFile{Bel2005}

\bibitem{BraDjiFay2006}
Brader L, Djibo H, Faye FG, Ghaout S, Lazar M, et~al. (2006) Towards a more
  effective response to desert locusts and their impacts on food security,
  livelihoods and poverty: Multilateral evaluation of the 2003--05 desert
  locust campaign.
\newblock Technical report, United Nations Food and Agriculture Organization.
\bibAnnoteFile{BraDjiFay2006}

\bibitem{SpeHunWat2008}
Speight MR, Hunter MD, Watt AD (2008) Ecology of Insects: Concepts and
  Applications.
\newblock Hoboken, NJ: Wiley-Blackwell, 2nd edition.
\bibAnnoteFile{SpeHunWat2008}

\bibitem{ColDesSim1998}
Collett M, Despland E, Simpson SJ, Krakauer DC (1998) Spatial scales of desert
  locust gregarization.
\newblock Proc Natl Acad Sci 95: 13052-13055.
\bibAnnoteFile{ColDesSim1998}

\bibitem{DesColSim2000}
Despland E, Collett M, Simpson S (2000) Small-scale processes in desert locust
  swarm formation: {H}ow vegetation patterns influence gregarization.
\newblock Oikos 88: 652-662.
\bibAnnoteFile{DesColSim2000}

\bibitem{AppHei1999}
Applebaum SW, Heifetz Y (1999) Density-dependent physiological phase in
  insects.
\newblock Ann Rev Entomol 44: 317-41.
\bibAnnoteFile{AppHei1999}

\bibitem{PenSim2009}
Pener MP, Simpson SJ (2009) Locust phase polyphenism: An update.
\newblock Adv Insect Physiol 36: 1 - 272.
\bibAnnoteFile{PenSim2009}

\bibitem{Dir1953}
Dirsh VM (1953) Morphometrical studies on phases of the desert locust
  (\textit{{S}chistocera {G}regaria} {F}orsk{\aa}l).
\newblock Anti-Locust Bull : 1-34.
\bibAnnoteFile{Dir1953}

\bibitem{IslRoeSim1994}
Islam MS, Roessingh P, Simpson SJ, McCaffery AR (1994) Parental effects on the
  behavior and coloration of nymphs of the desert locust {S}chistocera
  gregaria.
\newblock J Insect Physiol 40: 173-181.
\bibAnnoteFile{IslRoeSim1994}

\bibitem{SchAlb1999}
Schmidt GH, Albutz R (1999) Identification of solitary and gregarious
  populations of the desert locust, {S}chistocerca gregaria, by experimental
  breeding.
\newblock Entomol Gen 24: 161-175.
\bibAnnoteFile{SchAlb1999}

\bibitem{SimMcCHag1999}
Simpson SJ, McCaffery AR, H{\"{a}}gele BF (1999) A behavioral analysis of phase
  change in the desert locust.
\newblock Biol Rev 74: 461-480.
\bibAnnoteFile{SimMcCHag1999}

\bibitem{RogMatDes2003}
Rogers S, Matheson T, Despland E, Dodgson T, Burrows M, et~al. (2003)
  Mechanosensory-induced behavioural gregarization in the desert locust
  {S}chistocerca gregaria.
\newblock J Exp Bio 206: 3991-4002.
\bibAnnoteFile{RogMatDes2003}

\bibitem{SimDesHag2001}
Simpson SJ, Despland E, H{\"{a}}gele BF, Dodgson T (2001) Gregarious behavior
  in desert locusts is evoked by touching their back legs.
\newblock Proc Natl Acad Sci 98: 3895-3897.
\bibAnnoteFile{SimDesHag2001}

\bibitem{AnsRogOtt2009}
Anstey ML, Rogers SM, Ott SR, Burrows M, Simpson SJ (2009) Serotonin mediates
  behavioral gregarization underlying swarm formation in desert locusts.
\newblock Science 323: 627-630.
\bibAnnoteFile{AnsRogOtt2009}

\bibitem{RomCouSch2009}
Romanczuk P, Couzin ID, Schimansky-Geier L (2009) Collective motion due to
  individual escape and pursuit response.
\newblock Phys Rev Lett 102: 010602.
\bibAnnoteFile{RomCouSch2009}

\bibitem{BuhSumCou2006}
Buhl J, Sumpter DJ, Couzin ID, Hale JJ, Despland E, et~al. (2006) From disorder
  to order in marching locusts.
\newblock Science 312: 1402-1406.
\bibAnnoteFile{BuhSumCou2006}

\bibitem{Sum2010}
Sumpter DJ (2010) Collective Animal Behavior.
\newblock Princeton, NJ: Princeton University Press.
\bibAnnoteFile{Sum2010}

\bibitem{YatErbEsc2009}
Yates CA, Erban R, Escudero C, Couzin ID, Buhl J, et~al. (2009) Inherent noise
  can facilitate coherence in collective swarm motion.
\newblock Proc Natl Acad Sci 106: 5464-5469.
\bibAnnoteFile{YatErbEsc2009}

\bibitem{BazBuhHal2008}
Bazazi S, Buhl J, Hale JJ, Anstey ML, Sword GA (2008) Collective motion and
  cannibalism in locust migratory bands.
\newblock Curr Biol 18: 735-739.
\bibAnnoteFile{BazBuhHal2008}

\bibitem{BazRomTho2011}
Bazazi S, Romanczuk P, Thomas S, Schimansky-Geier L, Hale JJ, et~al. (2011)
  Nutritional state and collective motion: From individuals to mass migration.
\newblock Proc Roy Soc B 278: 356-363.
\bibAnnoteFile{BazRomTho2011}

\bibitem{VicCziBen1995}
Vicsek T, Czirok A, Ben~Jacob E, Cohen I, Shochet O (1995) Novel type of
  phase-transition in a system of self-driven particles.
\newblock Phys Rev Lett 75: 1226-1229.
\bibAnnoteFile{VicCziBen1995}

\bibitem{EdeWatGru1998}
Edelstein-Keshet L, Watmough J, Gr{\"{u}}nbaum D (1998) Do travelling band
  solutions describe cohesive swarms? {A}n investigation for migratory locusts.
\newblock J Math Bio 36: 515-549.
\bibAnnoteFile{EdeWatGru1998}

\bibitem{TopBerLog2008}
Topaz CM, Bernoff AJ, Logan S, Toolson W (2008) A model for rolling swarms of
  locusts.
\newblock Euro Phys J ST 157: 93--109.
\bibAnnoteFile{TopBerLog2008}

\bibitem{HolChe1996}
Holt J, Cheke R (1996) Models of desert locust phase changes.
\newblock Ecol Model 91: 131-137.
\bibAnnoteFile{HolChe1996}

\bibitem{ParMar2005}
Partan SR, Marler P (2005) Issues in the classification of multimodal
  communication signals.
\newblock Am Nat 166: 231-245.
\bibAnnoteFile{ParMar2005}

\bibitem{EftVriLew2007}
Eftimie R, de~Vries G, Lewis MA (2007) Complex spatial group patters result
  from different animal communication mechanisms.
\newblock Proc Natl Acad Sci 104: 6974-6979.
\bibAnnoteFile{EftVriLew2007}

\bibitem{KelSeg1971b}
Keller EF, Segel LA (1971) Model for chemotaxis.
\newblock J Theor Biol 30: 225-234.
\bibAnnoteFile{KelSeg1971b}

\bibitem{Oku1980}
Okubo A (1980) Diffusion and Ecological Problems.
\newblock New York: Springer.
\bibAnnoteFile{Oku1980}

\bibitem{MogEde1999}
Mogilner A, Edelstein-Keshet L (1999) A non-local model for a swarm.
\newblock J Math Bio 38: 534-570.
\bibAnnoteFile{MogEde1999}

\bibitem{TopBerLew2006}
Topaz CM, Bertozzi AL, Lewis MA (2006) A nonlocal continuum model for
  biological aggregation.
\newblock Bull Math Bio 68: 1601-1623.
\bibAnnoteFile{TopBerLew2006}

\bibitem{LevTopBer2009}
Leverentz AJ, Topaz CM, Bernoff AJ (2009) Asymptotic dynamics of
  attractive-repulsive swarms.
\newblock SIAM J Appl Dyn Sys 8: 880-908.
\bibAnnoteFile{LevTopBer2009}

\bibitem{BerTop2011}
Bernoff AJ, Topaz CM (2011) A primer of swarm equilibria.
\newblock SIAM J Appl Dyn Sys 10: 212-250.
\bibAnnoteFile{BerTop2011}

\bibitem{DOrChuBer2006}
D'Orsogna MR, Chuang YL, Bertozzi AL, Chayes L (2006) Self-propelled particles
  with soft-core interactions: Patterns, stability, and collapse.
\newblock Phys Rev Lett 96: 104302.1-104302.4.
\bibAnnoteFile{DOrChuBer2006}

\bibitem{GraSwoAns2009}
Gray LJ, Sword GA, Anstey ML, Clissold FJ, Simpson SJ (2009) Behavioural phase
  polyphenism in the {A}ustralian plague locust ({C}hortoicetes terminifera).
\newblock Biol Lett 5: 306-309.
\bibAnnoteFile{GraSwoAns2009}

\bibitem{DesSim2000b}
Despland E, Simpson S (2000) The role of food distribution and nutritional
  quality in behavioural phase change in the desert locust.
\newblock Animal Behaviour 59: 643-652.
\bibAnnoteFile{DesSim2000b}

\bibitem{DesSim2000}
Despland E, Simpson S (2000) Small-scale vegetation patterns in the parental
  environment influence the phase state of hatchlings of the desert locust.
\newblock Physiol Entomol 25: 74-81.
\bibAnnoteFile{DesSim2000}

\bibitem{MogEdeBen2003}
Mogilner A, Edelstein-Keshet L, Bent L, Spiros A (2003) Mutual interactions,
  potentials, and individual distance in a social aggregation.
\newblock J Math Bio 47: 353-389.
\bibAnnoteFile{MogEdeBen2003}

\bibitem{SwoLecSim2010}
Sword GA, Lecoq M, Simpson SJ (2010) Phase polyphenism and preventative locust
  management.
\newblock J Insect Physiol 56: 949-57.
\bibAnnoteFile{SwoLecSim2010}

\bibitem{BuhSwoCli2011}
Buhl J, Sword G, Clissold F, Simpson S (2011) Group structure in locust
  migratory bands.
\newblock Behav Ecol and Sociobiol 65: 265-273.
\bibAnnoteFile{BuhSwoCli2011}

\bibitem{Locustvideo}
Available: www.youtube.com/watch?v=1yny2r3hg2q. {A}ccessed 6 {J}uly 2012.
\bibAnnoteFile{Locustvideo}

\end{thebibliography}

\end{document}

% Title must be 150 characters or less
\begin{flushleft}
{\Large
\textbf{Supporting Information}
}
% Insert Author names, affiliations and corresponding author email.
\\
Chad M. Topaz$^{1,\ast}$, 
Maria R. D'Orsogna$^{2}$, 
Leah Edelstein-Keshet$^{3}$
Andrew J. Bernoff$^{4}$
\\
\bf{1} Department of Mathematics, Statistics, and Computer Science, Macalester College, Saint Paul, Minnesota, United States of America
\\
\bf{2} Department of Mathematics, California State University at Northridge, Los Angeles, California, United States of America
\\
\bf{3} Department of Mathematics, University of British Columbia, Vancouver, British Columbia, Canada
\\
\bf{4} Department of Mathematics, Harvey Mudd College, Claremont, California, United States of America
\\
$\ast$ E-mail: ctopaz@macalester.edu
\end{flushleft}

\section*{Model equations}

For convenience, we reintroduce our model equations.  Consider a two-dimensional domain $\Omega$ with spatial coordinate $\vec{x}=(x,y)$. Define $\rho(\vec{x},t)=s(\vec{x},t)+g(\vec{x},t)$ as the locust population density field, with $s(\vec x,t)$ and $g(\vec x, t)$ the solitary and gregarious components, respectively. The locust populations move with velocities $\vec{v}_{s,g}(\vec x, t)$ and obey the equations
\begin{subequations}
\label{eq:ge}
\begin{alignat}{4}
\dot{s} &+ \nabla \cdot (\vec{v}_s s)  &=  -f_2(\rho)s &+ f_1(\rho)g, \quad
\vec{v}_s = -\nabla(Q_s * \rho), \\
\label{eq:ge2}
\dot{g} &+ \nabla \cdot (\vec{v}_g g) &=  \phantom{-}f_2(\rho)s &-
f_1(\rho)g,  \quad \vec{v}_g = -\nabla(Q_g * \rho),
\end{alignat}
\end{subequations}
These equations generalize the classic swarming model
\begin{eqnarray}
\label{eq:social}
\rho_t + \nabla \cdot (\rho \vec{v}) = 0, \quad \vec{v} = -
\int_\Omega {\nabla Q}(\vec{x}- \vec{x}')  \rho (\vec{x}',t)
d\vec{x}',
\end{eqnarray}
which describes a single population density field advected by a velocity field arising from social interactions. Eq.\,\eqref{eq:social} has been studied extensively in one and two spatial dimensions for various social interaction functions represented by $Q$, whose negative gradient is the effective social force \cite{TopBer2004,BerLau2007,BodVel2005,BodVel2006}. Depending on $Q$, solutions include steady swarms, spreading populations, and contracting groups (\emph{i.e.}, blow-up) \cite{BerLau2007, LevTopBer2009, BerTop2011}.
  
In our two-phase model  Eqs.\,\eqref{eq:ge}, the velocities are
\begin{equation}
\label{eq:v}
 \vec{v}_{s,g} (\vec x, t)= -\nabla Q_{s,g} * \rho \equiv -
\int_\Omega {\nabla Q_{s,g}}(\vec{x} - \vec{x}')  \rho (\vec{x}',t) \,
d\vec{x}' ,
\end{equation}
and the social interaction potentials $Q_{s,g}$ are
\begin{equation}
\label{eq:Q}
Q_s(\vec{x}-\vec{x}') = R_s \exp{-|\vec{x}-\vec{x}'|/r_s}, \quad Q_g(\vec{x}-\vec{x}') = R_g
\exp{-|\vec{x}-\vec{x}'|/r_g} - A_g \exp{-|\vec{x}-\vec{x}'|/a_g}.
\end{equation}
Here, $R_s, R_g, A_g$ are interaction magnitudes and $r_s, r_g$ and $a_g$ are interaction length scales. We require $R_g a_g - A_g r_g > 0$ and $A_g a_g^2 - R_g r_g^2>0$ so that $Q_g$ includes short range repulsion and long range attraction, as in \cite{LevTopBer2009,BerTop2011, ChuDOrMar2007}, as this is the clumping regime, appropriate to capture the tendency of gregarious locusts to aggregate. We model the density-dependent rates of interconversion of the solitary and gregarious forms as
\begin{equation}
\label{eq:rates}
f_1(\rho) = \frac{\delta_1}{1+ \left( \rho/k_1 \right)^2}, \quad
f_2(\rho) = \frac{\delta_2 \left( \rho/k_2 \right)^2}{1+ \left(
  \rho/k_2 \right)^2}.
\end{equation}
The parameters $\delta_{1,2}$ are maximal rates and $k_{1,2}$ are characteristic locust densities at which the transitions occur at half
of their maximal values. To the best of our knowledge, our work is the first to consider locust phase changes via continuum  modeling of locust density \cite{BerLau2007, BodVel2005, BodVel2006, TopBer2004}.

\section*{Parameter selection and estimation}

As discussed in the main text, for our numerical results, we use two different sets of phase change parameters. For both sets, we use the same social interactions parameters, and we now describe our choices for these.

To estimate $R_s$, $R_g$, and $A_g$, we use explicit velocity computations. The speed of a locust when it is alone varies between $72\ \mbox{-}\ 216\ m/hr$, while the speed of a locust in a group varies in a tighter range of $144\ \mbox{-}\ 216\ m/hr$ \cite{BazRomTho2011}. To make a rough estimate of $R_s$, we imagine a hypothetical semi-infinite density field $\rho(x,y) = \rho_{group} \Heaviside(x)$ where $\Heaviside(x)$ is
the Heaviside function and, as mentioned in the main text, $\rho_{group} = 65\ locusts/m^2$ is the approximate critical density of a gregarious group~\cite{SimDesHag2001}. A solitary locust placed at the swarm's edge (at the origin) should move to the left with maximal velocity $v_s^{\rm max} = -216\ m/hr$. From Eqn.\,\eqref{eq:v},
\begin{equation}
v_s(0,0) =  \left\{-\nabla Q_s * \rho_{group}
\Heaviside(x)\right\}\big|_{(0,0)} = v_s^{\rm max},
\end{equation}
which we solve to find $R_s = 11.87\ m^3/(hr \cdot locust)$. Similarly, a gregarious locust at the origin should move to the right with maximal velocity $v_g^{\rm max} = 216\ m/hr$, so
\begin{equation}
v_g(0,0) = \left\{-\nabla Q_g * \rho_{group}
\Heaviside(x)\right\}\big|_{(0,0)} = v_g^{\rm max}.
\end{equation}
A gregarious locust placed to the left of the swarm at a distance equal to the attraction length scale $a_g = 0.14\ m$ should also move to the right, but with a slower velocity which we take to be the minimal velocity in a crowd, $v_g^{\rm min}=144\ m/hr$. Thus
\begin{equation}
v_g(-0.14,0) = \left\{-\nabla Q_g * \rho_{group}
\Heaviside(x)\right\}\big|_{(-0.14,0)} = v_g^{\rm min}.
\end{equation}
These two conditions determine $R_g = 5.13\ m^3/(hr \cdot locust)$ and $A_g = 13.33\ m^3/(hr \cdot locust)$ In the main text, we present numerical simulations of Eqs.\,\eqref{eq:ge} in one spatial dimension. For these simulations, we take $\delta_{1,2}$, $r_s$, $r_g$, and $a_g$ as above, since these parameters do not depend on spatial dimension. For the remaining parameters, we follow a process
similar to that described above, and choose $k_{1,2} = k = 8\ locusts/m$, $R_s = 6.83\ m^2/(hr \cdot locust)$, $R_g = 6.04\ m^2/(hr \cdot locust)$, and $A_g = 12.9\ m^2/(hr \cdot locust)$.

\section*{Homogeneous steady states}

For any set of initial conditions, the mean locust density $\rho_0$ is known, and corresponds to the total density at the homogeneous steady state (HSS). Accordingly, there is a family of homogeneous steady states parameterized by $\rho_0$. The corresponding solitary and gregarious HSS components, obtained by setting time and space derivatives to zero in Eqs.\,\eqref{eq:ge} are
\begin{subequations}
\label{eq:steadystate}
\begin{eqnarray}
s_0 & = & \frac{\rho_0 \delta_1 k_1^2 (k_2^2+\rho_0^2)}{\delta_1 k_1^2
k_2^2 + \delta_1 k_1^2 \rho_0^2 + \delta_2 k_1^2 \rho_0^2 +\delta_2
\rho_0^4} \label{eq:s0}, 
\\ g_0 & = & \frac{\delta_2 \rho_0^3
(k_1^2+\rho_0^2)}{\delta_1 k_1^2 k_2^2 + \delta_1 k_1^2 \rho_0^2 +
\delta_2 k_1^2 \rho_0^2 +\delta_2 \rho_0^4} \label{eq:g0}.
\end{eqnarray}
\end{subequations}
When we later consider stability of homogeneous steady states, it will be convenient to discuss the fractions $\phi_{s,g}$ of solitarious and gregarious locusts, where $\phi_s + \phi_g = 1$. Using Eqn.\,\eqref{eq:steadystate}, we know that for homogeneous steady states,
\begin{subequations}
\label{eq:phigrho}
\begin{eqnarray}
\phi_g & = & \frac{g_0}{s_0+g_0},\\
&  = & \frac{1}{s_0/g_0+1},\\
& = & \biggl\{1 + \gamma K^2 \frac{1+\psi^2}{\psi^2(\psi^2+K^2)}\biggr\}^{-1}.
\end{eqnarray}
\end{subequations}
Here, $\gamma = \delta_1/\delta_2$ is the ratio of maximal solitarization rate to maximal gregarization rate,  $K = k_1/k_2$ is the ratio of the characteristic solitarization and gregarization densities for individuals, and $\psi = \rho_0/k_2$ is a rescaled density. Note that $\phi_g$ is monotonically increasing in $\psi$, and hence in $\rho_0$; that is to say, as total density increases, the gregarious fraction increases.

\section*{Linear stability analysis}

To study the stability of the HSS in Eqs.\,\eqref{eq:steadystate}, we consider small perturbations  $s_1, g_1$ about $s_0, g_0$
\begin{equation}
\label{eq:pert1}
s(\vec{x},t) = s_0+s_1(\vec{x},t), \quad g(\vec{x},t) = g_0+g_1(\vec{x},t),
\end{equation}
\noindent
so that $\rho(\vec{x},t) =s_0+g_0+ s_1(\vec{x},t)+g_1(\vec{x},t)$. Substituting Eqn.\,\eqref{eq:pert1} into Eqn.\,\eqref{eq:ge} and expanding to first order in the perturbations, we find the linearized equations
\begin{subequations}
\label{eq:linearized}
\begin{eqnarray}
\dot{s}_1 & = & s_0 Q_s * \nabla^2 (s_1+g_1)  - A s_1 + B g_1, \\
\dot{g}_1 & = & g_0 Q_g * \nabla^2 (s_1+g_1) + A s_1 - B g_1,
\label{eq:fourier}
\end{eqnarray}
\end{subequations}
where
\begin{subequations}
\begin{eqnarray}
A & = & f_2(\rho_0)+f_2^\prime(\rho_0)s_0-f_1^\prime(\rho_0)g_0,\\
B & = & f_1(\rho_0)+f_1^\prime(\rho_0)g_0-f_2^\prime(\rho_0)s_0.
\end{eqnarray}
\end{subequations}
Here, $A,B > 0$ for all $\rho_0 > 0$ since $f_1$ is a monotonically increasing function of $\rho_0$ and $f_2$ is a monotonically decreasing one. To further analyze the linearized equations, we Fourier expand the perturbations as
\begin{equation}
\label{eq:perturb}
s_1(\vec{x},t) = \sum_\vec{q} \mathcal{S}_\vec{q}(t) \exp{i \vec{q}
  \cdot \vec{x}}, 
\quad s_2(\vec{x},t) = \sum_\vec{q} \mathcal{G}_\vec{q} (t) \exp{i
  \vec{q} \cdot
\vec{x}}.
\end{equation}
We allow for an infinitely large domain so that there are no restrictions on $\vec q$; in other situations, $\vec{q}$ must be suitably restricted in order to satisfy boundary conditions. Substituting Eqn.\,\eqref{eq:perturb} into Eqn.\,\eqref{eq:linearized} yields ordinary differential equations for each Fourier mode amplitude.  We write these in matrix form,
\begin{subequations}
\begin{gather}
\frac{d}{dt}
\begin{pmatrix}
\mathcal{S}_q \\\mathcal{G}_q
\end{pmatrix}
=
\mat{L}(q)
\begin{pmatrix}
\mathcal{S}_q \\ \mathcal{G}_q
\end{pmatrix}, \\
\mat{L}(q) \equiv \begin{pmatrix} -s_0 q^2 \wh{Q}_s(q) - A & -s_0 q^2
  \wh{Q}_s(q)+B \\ -g_0 q^2 \wh{Q}_g(q) + A & -g_0 q^2 \wh{Q}_g(q) - B
\end{pmatrix}.
\end{gather}
\end{subequations}
Here, $q = |\vec{q}|$ is the perturbation wavenumber, and $\wh{Q}_{s,g}(q)$ are the Fourier transforms of the two dimensional social interaction potentials,
\begin{eqnarray}
\label{eq:qhats}
\widehat{Q}_s(q)& =&  \frac{2 \pi R_s r_s^2}{(1+r_s^2 q^2)^{3/2}}, \\
\widehat{Q}_g(q) & = &   \frac{2 \pi R_g r_g^2}{(1+r_g^2q^2)^{3/2}}-
\frac{2 \pi A_g a_g^2}{(1+a_g^2 q^2)^{3/2}}.
\end{eqnarray}

The eigenvalues $\lambda_{1,2}(q)$ of $\mat{L}(q)$ are
\begin{equation}
\lambda_1(q) = -q^2 \left[s_0 \wh{Q}_s(q) + 
g_0 \wh{Q}_g(q)\right], \quad \lambda_2 = -(A+B).
\end{equation}
Since $\lambda_2 < 0$, instability occurs only when $\lambda_1 > 0$. For convenience, we rewrite $\lambda_1$ in terms of the gregarious mass fraction $\phi_g$,
\begin{equation}
\label{eq:eigrewrite}
\lambda_1(q) = - \rho_0 q^2 \left[(1-\phi_g) \wh{Q}_s(q) + 
\phi_g \wh{Q}_g(q)\right].
\end{equation}
Now we factor out the attractive part of the gregarious term, namely
\begin{equation}
\phi_g\frac{2\pi A_g a_g^2}{(1+a_g^2 q^2)^{3/2}}.
\end{equation}
This yields
\begin{equation}
\label{eq:eigrewrite2}
\lambda_1(q) = - \rho_0 q^2 \phi_g\frac{2\pi A_g a_g^2}{(1+a_g^2 q^2)^{3/2}} \biggl[ \frac{1 - \phi_g}{\phi_g} \frac{R_s r_s^2}{A_g a_g^2}\frac{(1+a_g^2 q^2)^{3/2}}{(1+r_s^2 q^2)^{3/2}} + \frac{R_g r_g^2}{A_g a_g^2}\frac{(1+a_g^2 q^2)^{3/2}}{(1+r_g^2 q^2)^{3/2}} -  1 \biggr].
\end{equation}
Since the prefactor is negative, and we seek conditions for a positive eigenvalue (signifying growth of perturbations, and hence instability), we focus on when the term in square brackets becomes negative. The dependence on $\phi_g$ occurs via the prefactor $(1 - \phi_g)/\phi_g$ in front of a positive term.  For possible instability, this term should be small, meaning that $\phi_g$ should be sufficiently large (since this  prefactor is monotonically decreasing with $\phi_g$). Since $\phi_g$ increases monotonically with $\rho_0$ (as discussed above), instability may occur as $\rho_0$ is increased.

We now show that instability first occurs at the wavenumber $q=0$ (meaning that perturbations that first lead to instability are long wavelength). We again focus on the bracketed quantity in Eq.\,\eqref{eq:eigrewrite2}. If this term becomes negative, it must do so for the value of $q$ at which the first two terms are (together) minimized, since these are positive terms and the negative term, $-1$, is a constant. It is biologically reasonable to assume that  $a_g \geq r_s$ (with equality achieved for our chosen social interaction parameters). Therefore, the first term is either constant or monotonically increasing in $q$. It is also biologically reasonable to assume that  $a_g > r_g$, in which case the second term is monotonically increasing in $q$. Thus, the first two terms together are monotonically increasing in $q$, so their minimum occurs at $q=0$, and this will be the first wavenumber to trigger instability. Thus, if we are looking for the instability that occurs as $\phi_g$ increases, it is sufficient to consider what happens at $q=0$.

We substitute $q=0$ into the bracketed term in Eqn.\,\eqref{eq:eigrewrite2} and ask for what value of $\phi_g$ the resultant expression changes sign (to find the threshold level of gregarious locust fraction needed for instability). Setting that bracketed term to zero we obtain
\begin{equation}
\phi_g^* = \frac{R_s r_s^2}{R_s r_s^2 - R_g r_g^2 + A_g a_g^2}.
\end{equation}
Instability is achieved for values of $\phi_g$ greater than this threshold value.

To obtain a more explicit condition for instability in terms of the density $\rho_0$, we substitute $\phi_g^*$ into Eq.\,\eqref{eq:phigrho}, which relates gregarious fraction to total (scaled) density. Rearranging, we obtain the biquadratic equation
\begin{equation}
A \psi^4 + B \psi^2 + C = 0,
\end{equation}
where
\begin{subequations}
\begin{eqnarray}
A & = & \frac{1}{\phi_g^*} - 1,\\
B & = & K^2 \biggl(\frac{1}{\phi_g^*} - 1 - \gamma\biggr),\\
C & = & -\gamma K^2.
\end{eqnarray}
\end{subequations}
For any biologically meaningful solutions, the solution for $\psi^2$ must be positive. From the quadratic formula, we have
\begin{equation}
\psi^2 = \frac{-B \pm \sqrt{B^2 - 4AC}}{2A}.
\end{equation}
Since $A>0$ and $C<0$, the discriminant is positive. Hence, for the plus sign choice, $\psi^2 > 0$. For the minus sign choice, $\psi^2 < 0$ and hence we eliminate this possibility. The final result for the critical scaled density is
\begin{equation}
\psi^* = \sqrt{\frac{-B + \sqrt{B^2 - 4AC}}{2A}}.
\end{equation}
This is the result that we use to produce instability contours in the $K$-$\gamma$ plane (Fig. 2 in the main paper).

\section*{Numerical simulation method}

We simulate  Eqs.\,\eqref{eq:ge}-\eqref{eq:rates} in one spatial dimension.  We use periodic boundary conditions on a domain of length $L$ with a fine grid consisting of $N=1024$ points (necessary to resolve the steep edges of clusters that form). To approximate an unbounded domain, one may take the limit of large $L$. The social interactions $Q_{s,g}$ in \eqref{eq:Q} must be adapted to be commensurate with a periodic domain. We begin with the function $Q(x) = \exp{-|x|/r}$, which is the building block of $Q_{s,g}$. We
calculate the discrete Fourier transform $\mathcal{F}$ of $-\partial_x Q$ on our domain as
\begin{equation}
\label{eq:dft}
\mathcal{F} \{-\partial_x Q(x)\} = -\frac{i}{r} \frac{\Delta
\sin(\Delta q)}{\cosh(\Delta/r)-\cos(\Delta q)},
\end{equation}
where $r$ is the decay length scale in $Q$ and $\Delta = L/N$ is the grid spacing. From Eqn.\,\eqref{eq:dft} it is straightforward to
compute the Fourier transforms of $Q_{s,g}$.  Convolutions are equivalent to products in Fourier space, providing excellent computational savings (and thus justifying the choice of a periodic domain). We compute  velocities by convoluting the density with $-\partial_x Q_{s,g}$ pseudospectrally. The flux term in Eqs.\,\eqref{eq:ge} is instead  evaluated via a fourth-order accurate central finite difference.

The emergence of discontinuities in $s$ and $g$  causes ringing in the pseudospectral evaluation of the velocity term. In order to smooth this effect, we incorporate small amounts of numerical diffusion. Another standard approach would be to incorporate high wave
number filtering in the simulation. We choose numerical diffusion because it also serves as the macroscopic description of random motion, which locusts certainly display. We implement diffusion in a split-step manner, alternating with the dynamics of Eqs.~\eqref{eq:ge}-\eqref{eq:rates}. Time-stepping is performed with the fourth-order Runge-Kutta method. We also threshold our velocity
field at every time step so that it does not exceed $v_g^{\rm max}$. Without this thresholding, individual locusts achieve velocities of up to approximately 1.5 times $v_g^{\rm max}$ at an intermediate stage of our simulation. It is crucial to point out that this thresholding only affects the speed of the transient clumps; it does not affect the initial instability (which is small amplitude, and thus has a small velocity) and similarly, it does not affect the late-stage bulk dynamics (which are nearly spatially stationary).

% The bibtex filename
\bibliography{locustbib}